\documentclass[aps,floatfix,twocolumn,showpacs]{revtex4}
\usepackage{natbib}
\usepackage{dcolumn}
\usepackage{graphicx}
\newcommand{\sr}{{\rm sr}}

\newcommand{\mnras}{Mon.~Not.~R.~Astron.~Soc.}

\begin{document}
\title{Dynamical dark energy: Current constraints and forecasts}
\author{Amol Upadhye$^{1}$}
\author{Mustapha Ishak$^{2}$}
\author{Paul J. Steinhardt$^{1}$ }
\affiliation{
$^1$Department of Physics, Princeton University,
  Princeton, NJ 08544, USA \\
$^2$Department of Astrophysical Sciences, Princeton University,
  Princeton, NJ 08544, USA}
\date{\today}
\begin{abstract}
We consider how well the dark energy equation of state 
$w$ as a function of red shift $z$ will be measured using current and 
anticipated experiments.   We use a procedure
which takes fair account of the uncertainties in the functional
dependence of $w$ on $z$, as well as the parameter degeneracies, and 
avoids the use of strong prior constraints.  We apply the 
procedure to current data from WMAP, SDSS, and the supernova searches, and 
obtain results that are consistent with other analyses using different 
combinations of data sets.  The effects of systematic experimental errors
and variations in the analysis technique are discussed.  Next, we use the same
procedure to forecast the dark energy constraints achieveable by the 
end of the decade, assuming 8 years of WMAP data and
realistic projections for  ground-based measurements of supernovae and 
weak lensing.  We find the $2 \sigma$ 
constraints on the current value of $w$ to be $\Delta w_0 (2 \sigma ) = 0.20$, 
and on $dw/dz$ (between $z=0$ and $z=1$) to be  $\Delta w_1 (2 \sigma )=0.37$.
Finally, we compare these limits to other projections in the literature.
Most show only a modest improvement; others show a more substantial 
improvement, but there are serious concerns about systematics.
The remaining  uncertainty
still allows a significant span of competing dark energy models.  Most
likely, new kinds of measurements, or experiments more sophisticated 
than those currently planned, are needed to reveal the true nature 
of dark energy.
\end{abstract}


%
\maketitle
\section{Introduction}
\label{sec:intro}
Several years ago, observations of supernovae of type Ia (SNe Ia) demonstrated that the expansion of the universe is accelerating \cite{Riess98, Perlmutter98}. Associated with this acceleration is dark energy, a component of the universe with negative pressure, that makes up a significant fraction of the total energy density.  Recent years have seen the supernova evidence for dark energy continue to mount \cite{Riess00, Riess01, Tonry, Knop, Barris, Riess04}.  Meanwhile, cosmic microwave background (CMB) data from the WMAP project \cite{Bennett, Spergel}, in combination with information about either the Hubble constant \cite{HSTkey} or the galaxy power spectrum \cite{SDSS_pwrspec, SDSS_params}, also requires the existence of dark energy.  Cross-correlations between CMB anisotropies and matter power spectrum inhomogeneities provide evidence for a late-time integrated Sachs-Wolfe (ISW) effect \cite{Fosalba03, Fosalba04, Scranton, Niayesh1, Boughn, NikhilISW04}.  This ISW effect indicates a recent change in the inhomogeneous gravitational potential, providing further evidence for dark energy.

One way to characterize the dark energy is to measure its equation of state, the ratio $w \equiv P / \rho$ of its pressure $P$ and energy density $\rho$.  The simplest model of the dark energy is a cosmological constant $\Lambda$, which has a constant $w=-1$.  Quintessence models describe the dark energy as a dynamical scalar field with an equation of state $w\geq-1$, and in general, $w$ will not be constant in time \cite{Peebles, Ratra, Zlatev, Steinhardt-quintessence}.   Furthermore, models of extended quintessence or ``phantom energy'' even allow $w<-1$ \cite{CaldwellPhantom}.  Models of the universe incorporating a cosmological constant or quintessence, in addition to cold dark matter, are known as $\Lambda$CDM or QCDM models, respectively.

Discovering whether or not the dark energy is a cosmological constant is the primary goal of the study of dark energy.  If dark energy is shown not to be a cosmological constant, then the next important issues are whether or not $w<-1$, which can be theoretically problematic \cite{Carroll} (however, see \cite{onemli02, onemli04, onemli05, DoranJaeckel, DeDeo} for an alternative viewpoint), and whether or not $w$ is changing with time.   While a quintessence with constant $w\approx-1$ is very difficult to distinguish quantitatively from a cosmological constant, the qualitative difference for fundamental physics is enormous, so it is critical that the maximum effort be made to reduce the uncertainty in $|w+1|$ and its time derivative.  Ultimately, the precise quantitative values of $w$ and its time derivative are important for model building, but this is less crucial, at present, than the qualitative issue of whether the dark energy is dynamical or not.  In order to address such questions about the nature of the dark energy, it is crucial that the equation of state and its time variation be determined observationally.

It has been known for some time that the cosmological probes used for stuyding the dark energy are plagued by numerous parameter degeneracies \cite{Bond1994, Bond1997, Matias1997, White1999, Huey, cosmicConfusion, Maor2001}.  Because of these degeneracies, a large uncertainty in a cosmological parameter not directly related to the dark energy sector ({\em{e.g.}}, the matter density) can lead to a large uncertainty in the dark energy equation of state.  This sensitivity to other parameters means that one must incorporate the uncertainties in all parameters to obtain a realistic estimate of the uncertainties in the dark energy equation of state.  In particular, one should avoid the use of strong priors on the form of $w(z)$, the values of other parameters, and independent information coming from other experiments.  These can lead to underestimates of the uncertainty in $w$ by a very large factor.  We will also discuss how standard likelihood marginalization can give a misleading impression of the uncertainty in $w$.

The analysis presented here employs a $\chi^2$ minimization procedure in order to avoid possible problems with marginalization, and assumes only weak prior constraints.  We compare our $\chi^2$ minimization with the standard approach, the marginalization of a probability function computed using Monte Carlo Markov Chains.  It is argued that our procedure gives a more conservative assessment of constraints in certain degenerate parameter spaces.  In this sense, our procedure and the Markov Chain are complementary, as we will discuss.

We apply this $\chi^2$ minimization analysis to the currently available data.  The simultaneous determination of the dark energy equation of state and its red shift derivative is difficult without the combination of data from several cosmological probes.  Thus we analyze the latest data from WMAP, SN Ia searches, and the Sloan Digital Sky Survey.  Strong priors are avoided to prevent the underestimation of dark energy constraints.  Although the data favor $w(0)<-1$ and $w'(0)>0$, we are unable to rule out the cosmological constant conclusively.  Furthermore, we find the constraints on $w(z)$ to depend strongly on the parameterization chosen for $w(z)$, as well as on the location of the best-fit model in parameter space, despite the fact that we have restricted our study to two-parameter equations of state.  We also list possible biases in the cosmological probes, and discuss their effects on our results.  Our conclusion that the cosmological constant is consistent with current data agrees with previous analyses, as we discuss in Sec. \ref{sec:discussion}.  This establishes the validity of our analytic methods, which we can then apply with confidence to projections of future measurements.

Since the cosmological constant is consistent with current data, it is useful to ask precisely how well dark energy may be constrained in the future using known cosmological probes.  We address this question by applying our $\chi^2$ minimization procedure to simulations of the data from these probes.  Besides the CMB and SNe Ia, we simulate data from a probe of weak gravitational lensing (WL).   In the future, WL is expected to be useful for constraining dark energy \cite{Hu01,Huterer01,SNAP04,SongKnox,Albrecht,WangClusterCounts04,HutererTurner,Benabed,Abazajian,Refregier,Heavens,Simon,JainTaylor,BernsteinJain}.  Most of these studies have demonstrated that WL breaks parameter degeneracies in the CMB analysis,  and have shown the effectiveness of the combination of CMB and WL data in constraining dark energy.  

Our study is the first to use a joint analysis of simulated CMB, SN Ia, and WL data, rather than imposing prior constraints as a ``replacement'' for one of these data sets.  This is especially important for parameters such as $w(0)$ and $w'(0)$, which have strong degeneracies with other parameters.  Since our goal is the study of $w(z)$ constraints, rather than general parameter constraints, we assume the cosmological model outside of the dark energy sector to be as simple as possible.  Within this model, we are careful to avoid strong priors on the cosmological parameters.  It is well understood that each of these cosmological probes has systematic uncertainties associated with it, and that significant progress needs to be made in order for each one to reach the level of precision necessary for constraining dark energy.  Therefore, we attempt at every opportunity to strike a reasonable balance between optimism and realism.  

The paper is organized as follows.  Sec. \ref{subsec:DEModel} discusses several parameterizations of the dark energy equation of state.  Our choice of cosmological parameters, and the prior constraints imposed on them, are listed in Sec. \ref{subsec:cosmoParamDegenPriors}.  $\chi^2$ minimization is compared with marginalization in Sec. \ref{subsec:gridMin}.  Sec. \ref{subsec:currentChi2} summarizes our analysis of current data, and Secs. \ref{subsec:snMC}-\ref{subsec:wlMC} describe the simulation and analysis of future data.  Finally, Sec. \ref{sec:discussion} lists and interprets our findings, for our analysis of current data as well as our forecasts of future constraints, and Sec. \ref{sec:conclusions} discusses our conclusions.


\section{Analysis Methods}
\label{sec:analysis}
\subsection{Dark energy parameterization}
\label{subsec:DEModel}

In the spirit of reasonable optimism, we have assumed that the dark energy may be parameterized using an equation of state $w(z) = P / \rho$, where $P$ and $\rho$ are the pressure and dark energy density, respectively, and $w(z)$ is an unknown function of red shift.  We have also assumed that the dark energy sound speed is $c_s^{\phantom{s}2}=1$.  Without any theoretical guidance about the form of $w(z)$, the space of all possibilities is equivalent to an uncountably infinite set of cosmological parameters.  The data depend on $w(z)$ only through a multiple integral relation \cite{Maor2001}, so we lack the large number of measurements of $w(z)$ that would be necessary for non-parametric inference \cite{Jang03}.  Thus we must describe $w(z)$ using a small number of parameters in order to keep the analysis tractable.  The danger is that there is no ``natural'' way to parameterize $w(z)$, so the choice of parameterizations is essentially arbitrary.  Ideally, one would like to verify that any constraints obtained on $w(z)$ are parameterization-independent.  However, the space of all possible parameterizations is infinite-dimensional, so we're back to square one.  The only feasable option is to test for parameterization-independence by comparing constraints on a small number of ``reasonable-looking'' parameterizations.  

To begin with, we assume that $w(z)$ is an analytic function, and that it can be approximated at $z \ll 1$ by the first few terms in its Taylor series about $z=0$.  All known probes are sensitive to an integral of some function of $w(z)$, rather than to $w(z)$ itself, so any ``bumps and wiggles'' in $w(z)$ are effectively smoothed out.  It is reasonable to assume that this smoothed $w(z)$ can be approximated by a low-degree polynomial in $z$ out to $z \lesssim 1$.  From now on we proceed based on this assumption, with the caveat that the ``true'' equation of state can have added ``bumps and wiggles'' that disappear when smoothed.  (Note, however, that any large-scale kinks or sharp changes in the actual equation of state can be missed by such simple parameterizations.  For an alternative approach, see, {\em{e.g.}}, \cite{bassett04}.)

The simplest parameterizations useful for describing dynamical dark energy have at least two parameters, one describing the equation of state $w(0)$ at $z=0$ and the other parameterizing its red shift derivative $w'(0) \equiv \left. \frac{dw}{dz} \right|_{z=0}$.  A few parameterizations in the literature are
{\setlength\arraycolsep{2pt}
\begin{eqnarray}
w(z) & = & w_0 + w' z \quad \textrm{(Simple parameterization)} \label{eqn:EOSsimple}\\
w(z) & = & w_0 + \left( 1-a(z) \right) w_a \nonumber\\
     & = & w_0 + \frac{w_a z}{1+z} \quad \textrm{(Refs. \cite{polarski, Linder02})} \label{eqn:EOSlinder} \\
w(z) & = & w_0 + \left( 1-a(z) \right) w_a + \left( 1-a(z) \right)^2 w_b \nonumber\\
     & = & w_0 + \frac{w_a z}{(1+z)} + \frac{w_b z^2}{(1+z)^2} \quad \textrm{(Ref. \cite{UrosSdssLyaBias})} \label{eqn:EOSuros}. 
\end{eqnarray}}(Note that $w_a$ in (\ref{eqn:EOSuros}) differs by a factor of $-1$ from the definition introduced in \cite{UrosSdssLyaBias}.  We chose (\ref{eqn:EOSuros}) in order to facilitate comparison with (\ref{eqn:EOSlinder}) and (\ref{eqn:EOSparam}).)  We are interested in the simplest two-parameter equations of state, and as we will see, (\ref{eqn:EOSsimple}) is of limited utility for studying high red shift data.  This leaves only Eq. (\ref{eqn:EOSlinder}) out of the three parameterizations above.  Since we would like to compare two ``reasonable'' two-parameter equations of state, we introduce a fourth parameterization,
\begin{eqnarray}
\label{eqn:Q_vs_z1}
w(z)         &  = & \left\{ \begin{array}{cl}
w_0 + w_1 z      &       \textrm{if $z<1$} \\
w_0 + w_1        & \textrm{if $z \geq 1$}.
\end{array} \right. 
\label{eqn:EOSparam}
\end{eqnarray}
Dark energy constraints based on (\ref{eqn:EOSparam}) and (\ref{eqn:EOSlinder}) may be compared in order to estimate the parameterization-dependence of our results.  Table \ref{tab:EOSparam} compares all four parameterizations using their present-day derivatives $w'(0)$, their mean low red shift ($0<z<1$) derivatives $\bar{w}' \equiv w(1)-w(0)$, and their high red shift limits $\lim_{z \rightarrow \infty} w(z)$.  

\begin{table}[tb]  
\begin{center}
\begin{tabular}{|c||c|c|c|}
\hline
$w(z)$                 & $w'(0)$                  & $\bar{w}' = $          & $w(z)$ \\
                       &                          & $w(1)-w(0)$            & at $z \gg 1$ \\
\hline
\hline
$w_0 + w'z$            & $w'$                     & $w'$                   & divergent \\
\hline
$w_0 + \frac{w_a z}{(1+z)}$ 
                       & $w_a$                    & $\frac{1}{2}w_a$       & $w_0+w_a$ \\
\hline
$w_0 + \frac{w_a z}{(1+z)} + \frac{w_b z^2}{(1+z)^2}$ 
                       & $w_a$                    & $\frac{1}{2}w_a + \frac{1}{4}w_b$ 
                                                                           & $w_0+w_a+w_b$ \\

\hline
$w_0+w_1z$ if $z<1$    & $w_1$                    & $w_1$                  & $w_0+w_1$ \\
$w_0+w_1$ if $z\geq 1$ &                          &                        & \\
\hline
\end{tabular} 
\end{center}
\caption{Summary of the dark energy equation of state parameterizations used in the literature. \label{tab:EOSparam}}
\end{table}

The high red shift limit of each parameterization becomes important when considering data from the CMB.  Rather than providing information about $w(z)$ directly, the CMB data require only that the dark energy density be less than $\sim 10 \%$ of the critical density at the red shift $z_{dec} \approx 1100$ of photon decoupling \cite{CaldEtAl, CaldDoran}.  If $w(z)>0$, the dark energy density will increase faster with red shift than the matter density, making the CMB constraint extremely difficult to satisfy.  For the four dark energy parameterizations discussed above, the CMB effectively requires that $w(z) \leq 0$ at red shifts $z \gg 1$, that is, that the quantities listed in the fourth column of Table \ref{tab:EOSparam} are no greater than zero.

In each of the the two- and three-parameter equations of state (\ref{eqn:EOSsimple}, \ref{eqn:EOSlinder}, \ref{eqn:EOSuros}, \ref{eqn:EOSparam}), this high red shift constraint prevents $w(z)$ from changing too rapidly at low red shifts.  For fixed $w_0$, we see that: $w' \leq 0$ in (\ref{eqn:EOSsimple}); $w_1 \leq -w_0$ in (\ref{eqn:EOSparam}); and $w_a \leq -w_0$ in (\ref{eqn:EOSlinder}).  In particular, $w_1$ and $w_a$ have the same upper bound.  Note that these bounds arise purely from the choice of parameterizations for $w(z)$.  There is no fundamental physical reason for assuming that an equation of state which becomes positive at red shift $z=1$ will stay that way long enough to interfere with the physics at decoupling.

\begin{figure}[tb]
\includegraphics[width=2.1in, angle=270]{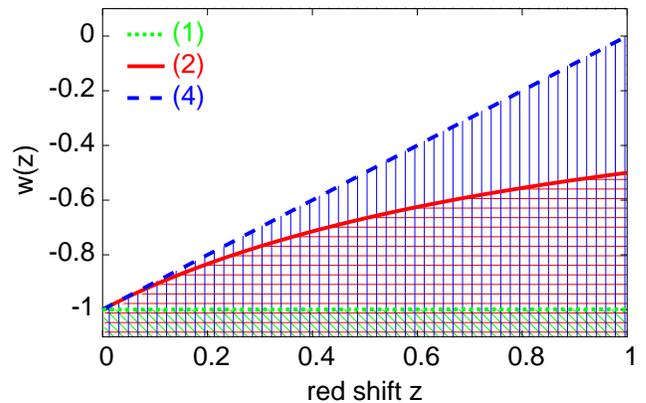}
\caption{Upper bounds on $w(z)$ for the three equations of state (\ref{eqn:EOSparam}), (\ref{eqn:EOSsimple}), and (\ref{eqn:EOSlinder}).  In all three cases, $w_0$ has been fixed at $-1$.}
\label{fig:wParams}
\end{figure}

Fig. \ref{fig:wParams} shows the upper bounds on the function $w(z)$ for each of the three parameterizations (\ref{eqn:EOSsimple}), (\ref{eqn:EOSlinder}), and (\ref{eqn:EOSparam}), assuming $w_0=-1$ in each case.  Parametrization (\ref{eqn:EOSsimple}) disallows essentially all positive $w'$; we will not study it further.  Meanwhile, (\ref{eqn:EOSlinder}) allows $w(z)$ to increase by ${\bar{w}}'=w(1)-w(0)=0.5$ at low $z$, while (\ref{eqn:EOSparam}) allows $w(z)$ to increase by ${\bar{w}}'=1$.  Some parameter combinations such as ($w_0=-1$, $\bar{w}'=0.6$) are acceptable in (\ref{eqn:EOSparam}), but are simply not allowed in (\ref{eqn:EOSlinder}).  That is, the two parameterizations cover different regions in the ($w$, $z$) plane.  Of course, these parameterization-dependent effects of the high red shift $w(z)$ constraint will be important only if the dark energy is found to have $w(z) \sim 0$ at $z \gg 1$.  

We choose to use parameterization (\ref{eqn:EOSparam}) for most of the subsequent work, since it allows the widest range of ${\bar{w}}'$ values.  Equation (\ref{eqn:EOSlinder}) is also studied, and the results based on the two parameterizations are compared in order to estimate the parameterization dependence of our dark energy constraints.  We describe the dynamics of the dark energy in terms of the ratio of the dark energy density to its value today, 
\begin{equation}
\mathcal{Q}(z) \equiv \rho_{de}(z) / \rho_{de}(0).
\label{eqn:rho_de_vs_z}
\end{equation}
For parameterization (\ref{eqn:EOSparam}), the fluid continuity equation can be used to show that
\begin{equation}
\mathcal{Q}(z)  \equiv  \left\{ \begin{array}{cl}
(1+z)^{3(1+w_0-w_1)} \; e^{3w_1z}      &       \textrm{if $z<1$,} \\
(1+z)^{3(1+w_0+w_1)} \; e^{3w_1(1-2 \ln 2)} & \textrm{if $z \geq 1$}.
\end{array} \right. 
\label{eqn:Q_vs_z}
\end{equation}
(A corresponding expression for equation of state (\ref{eqn:EOSlinder}) can be found in \cite{polarski, Linder02}.)  Given our assumption of a flat universe, the Hubble parameter $H(z) \equiv {\dot{a}}/a$ evolves with red shift as follows.
\begin{equation}
H(z) = H_0 \sqrt{\Omega_m(1+z)^3 + (1-\Omega_m)\mathcal{Q}(z)}
\label{eqn:H_vs_z}
\end{equation}
These results will be used in our discussions of the cosmological probes.

\subsection{Cosmological parameters, degeneracies, and priors}
\label{subsec:cosmoParamDegenPriors}

Cosmology may be described using a large number of parameters corresponding to a wide range of possible effects.  The approach taken by, {\em{e.g.}}, \cite{SDSS_params} is to test for as many of these effects as possible; they describe cosmology in terms of $13$ parameters.  On the other hand, our aim is more specific.  We wish to determine whether dynamical dark energy will be distinguishable from a cosmological constant by the end of the decade.  If we can answer this question in the negative after considering only a subset of these $13$ parameters, then the addition of more parameters will not change our conclusions.  We find that even with a relatively limited set of parameters we will not be able to rule out dynamical dark energy.  Thus, in the interests of simplicity and computational efficiency, we restrict ourselves to nine parameters.

Outside of the dark energy sector, we choose the simplest possible description of the universe that is consistent with observations.  The universe is assumed to be flat, and we do not include tensor modes, massive neutrinos, or primordial isocurvature perturbations.  Besides the dark energy parameters $w_0$ and $w_1$, our cosmological models are parameterized using: $h \equiv H_0 / $($100$ km sec${}^{-1}$ Mpc${}^{-1}$), where $H_0$ is the Hubble constant;  $\omega_m \equiv \Omega_m h^2 = \rho_m h^2 / \rho_{crit}$; $\omega_b \equiv \Omega_b h^2 = \rho_b h^2 / \rho_{crit}$; $\tau$, the optical depth to reionization; $A$, the normalization of the CMB power spectrum; $n_s$, the spectral index of the primordial power spectrum; $z_s$, the characteristic weak lensing source red shift.  (The WL source red shift was shown to be important in, {\em{e.g.}}, \cite{Ishak03, Tereno04,Ishak04b}.)  Actually, \cite{SDSS_params} shows that $n_s$ is not absolutely necessary, since the simple Harrison-Peebles-Zeldovich spectrum ($n_s=1$) fits the data well.  However, $n_s$ is a well-motivated parameter whose degeneracies with the dark energy parameters could be important.  Borrowing the terminology of \cite{SDSS_params}, we call ($h$, $\omega_m$, $\omega_b$, $\tau$, $n_s$, $A$) the ``vanilla'' parameters.  Thus, our analysis adds to the simple ``vanilla'' model one parameter describing weak lensing sources, and two parameters describing the dark energy.

Our constraints based on this limited set of parameters will be optimistic, since including more parameters will tend to increase the uncertainties in $w_0$ and $w_1$.  This is especially true when the parameters that we neglect are highly correlated with some of the parameters that we do consider.  Ref. \cite{SDSS_params} shows a strong degeneracy between the Hubble parameter $h$ and the curvature $\Omega_K$ that could potentially affect our dark energy constraints.  Also, \cite{Huterer01} points out that neutrinos become important when using weak lensing in the nonlinear regime to study dark energy.

\begin{figure}[tbh]
\begin{center}
\includegraphics[width=2.0in, angle=270]{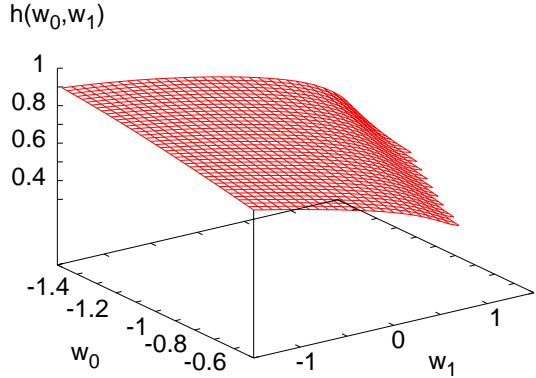}
\caption{$h(w_0,w_1)$ satisfying the CMB angular diameter distance degeneracy, with parameters $\omega_m=0.14$, $z_{dec}=1089$, and $d_{AC}^{(dec)}=14.0\;Gpc$ fixed to their WMAP \cite{Bennett} values.}
\label{fig:geomDeg}
\end{center}
\end{figure}

Even in our relatively simple parameter space, there exist several parameter degeneracies.  That is, to each point in parameter space there corresponds a degenerate region of observationally indistinguishable points.  A degeneracy that is particularly important to the study of dark energy is the angular diameter distance degeneracy \cite{Bond1997, White1999, Huey, cosmicConfusion}.  For flat universes containing matter and a dark energy of the form (\ref{eqn:EOSparam}), this implies that models with the same values of $\omega_m$, $\omega_b$, and the comoving angular diameter distance $d_{AC}^{(dec)}$ to the decoupling surface, will produce indistinguishable CMB power spectra.  $d_{AC}^{(dec)}$ as a function of $h$, $w_0$, and $w_1$ is given by
\begin{equation}
d_{AC}^{(dec)}(h,w_0,w_1) = c \int_0^{z_{dec}} \frac{dz'}{H(z')} ,
\label{eqn:d_AC}
\end{equation}
where $H(z')$ is given by (\ref{eqn:H_vs_z}).  Even when $\omega_m$ is fixed, each choice of $w_0$ and $w_1$ has a corresponding $h$ for which this degeneracy is satisfied, as shown in Fig. \ref{fig:geomDeg}.  Other important degeneracies include the large-scale CMB degeneracy \cite{Bond1994} among $\omega_m$, $\tau$, $n_s$, and $w(z)$; the small-scale CMB degeneracy \cite{Matias1997} among $\tau$, $A$, and $n_s$;  and the supernova luminosity distance degeneracy \cite{Maor2001} between $\Omega_m$ and $w(z)$.  The presence of these parameter degeneracies means that we must combine several cosmological probes in order to obtain reliable constraints.  Furthermore, combinations of known probes contain residual degeneracies, which we must deal with carefully in order to determine constraints on the dark energy.

\begin{table}[tb]  
\begin{center}
\begin{tabular}{|l||c|c|}
\hline
Parameter  & Lower Bound & Upper Bound \\
\hline
\hline
$h$        & $0.4$       & $1.1$ \\
\hline
$\omega_b$ & $0.003$      & $\omega_m$ \\
\hline
$\omega_m$ & $\omega_b$  & $1$ \\
\hline
$\Omega_m$ & $0$         & $1$ \\
\hline
$\Omega_K$ & $0$         & $0$ \\
\hline
$\tau$     & $0$         & $1$ \\
\hline
$A$        & $0.5$       & $1.5$ \\
\hline
$n_s$      & $0.5$       & $1.5$ \\
\hline
$z_s$      & $0$         & $1.5$ \\
\hline
$w_0$      & N/A         & $0$ \\
\hline
$w_0+w_1$  & N/A         & $0$ \\
\hline
\end{tabular} 
\end{center}
\caption{Summary of the prior constraints assumed for the $\chi^2$ minimization analysis. \label{tab:priors}}
\end{table}

In order to make a fair assessment of parameter degeneracies, it is necessary to avoid strong prior constraints on the cosmological parameters.  We have been careful to use very weak priors.  The dark energy equation of state is required to be negative or zero at all red shifts, $w(z) \leq 0$.  For the model (\ref{eqn:EOSparam}), this implies the two prior constraints $w_0 \leq 0$ and $w_0 + w_1 \leq 0$.  For physical reasons, we must assume $0<\Omega_m<1$, $\omega_b<\omega_m$, $\tau>0$, and $z_s>0$.  Besides these, the constraints $0.4<h<1.1$, $0.003<\omega_b$, $\omega_m<1$, $\tau<1$, $0.5<A<1.5$, $0.5<n_s<1.5$, and $z_s<1.5$ are imposed for computational convenience.  The above priors are summarized in Table \ref{tab:priors}.

The simulated data sets used here for constraint forecasts assumed a fiducial model \{$w_0=-1$, $w_1=0$, $h=0.7$, $\omega_m=0.15$, $\omega_b=0.023$, $\tau=0.1$, $A=0.8$, $n_s=1$, $z_s=1$\}.  This roughly corresponds to the ``vanilla lite'' $\Lambda$CDM model of \cite{SDSS_params}, with lensing sources assumed to be at characteristic red shift $1$.  Our chosen fiducial model implies $\Omega_m=0.31$, $\Omega_b=0.047$, and $\sigma_8=0.92$.  


\subsection{$\chi^2$ minimization}
\label{subsec:gridMin}

In order to provide an accurate picture of the dark energy constraints in a degenerate parameter space, a $\chi^2$ minimization procedure was used.  Constraints on the dark energy parameters were determined using the $1 \sigma$ and $2 \sigma$ contours of the $\chi^2$ function in the ($w_0$, $w_1$) plane.  At any given point ($w_0$, $w_1$), $\chi^2(w_0,w_1)$ was computed by minimizing over all other cosmological parameters.  

\begin{figure}[tb]
\includegraphics[width=5in, angle=270]{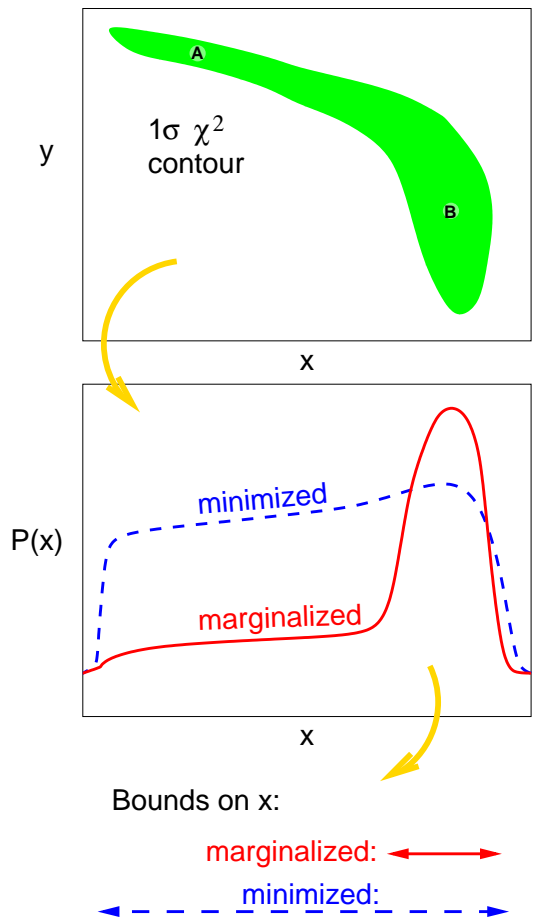}
\caption{Comparison of minimization and marginalization in a (hypothetical) degenerate parameter space.  Two models, $A$ and $B$, are labelled.
({\em{top}}): Sample $1 \sigma$ contour of $\chi^2(x,y)$ in a degenerate parameter space. 
({\em{mid}}): Probability functions based on minimization and marginalization.  The probability based on minimization (dashed line) is ${\mathcal{P}}_{min}(x) \propto exp \left( \min_{y} \chi^2(x,y) \right)$.  The marginalized probability (solid line) is ${\mathcal{P}}_{marg}(x) \propto \int exp \left( \chi^2(x,y) \right) {\mathcal{P}}_{prior}(x,y) dy$. 
({\em{bot}}): Constraints on $x$ based on minimized and marginalized probability functions.}
\label{fig:minVsMarg}
\end{figure}

$\chi^2$ minimization handles certain types of degeneracies more carefully than marginalization, the procedure most commonly used to obtain cosmological constraints, meaning that the two methods are complementary.  Marginalization favors models that fit well over ``large'' regions of parameter space.  Given two cosmological models with the same low $\chi^2$, marginalization will give extra weight to the model that sits in a large region of low-$\chi^2$ models.  Meanwhile, $\chi^2$ minimization will pick the best model while completely ignoring its surroundings.  These differences are illustrated in Fig. \ref{fig:minVsMarg}, based on a sample $\chi^2$ function of two arbitrary parameters $x$ and $y$.  The $1 \sigma$ contour of $\chi^2(x,y)$ is shown in Fig. \ref{fig:minVsMarg}(top).  Note that the contour has a small slope at low $x$, but is nearly parallel to the $y$ axis at high x.  

If we are interested only in constraints on $x$, then we can either minimize or marginalize over $y$.  Fig. \ref{fig:minVsMarg}(mid) sketches the probability functions obtained using the two different methods.  Consider the two models $A$ and $B$, at $x_A$ and $x_B$ respectively, whose $y$ values are chosen to minimize $\chi^2(x,y)$ at the corresponding $x$ values.  Minimization takes the viewpoint that, since $\chi^2_A \approx \chi^2_B$, the $x$ values $x_A$ and $x_B$ are approximately equally probable.  On the other hand, marginalization assigns a much greater probability to $x_B$, since there are many more low-$\chi^2$ models at $x_B$ than at $x_A$.  Thus the $\chi^2$-minimized probability distribution looks like a broad plateau, while the marginalized probability looks like a sharp peak with a ``shoulder.''  The bounds on $x$ obtained from these two methods, as shown in Fig. \ref{fig:minVsMarg}(bot), are quite different.  Note that this discussion is not just academic.  The CMB $95\%$ probability contour in the ($h$, $\Omega_{tot}$) plane, shown in Fig. 7 of \cite{SDSS_params}, is qualitatively very similar to the contour shown here in Fig. \ref{fig:minVsMarg}(top).

The parameter bounds derived from marginalization have excluded models such as $A$, which are within the $1 \sigma$ $\chi^2$ contour, simply because the low-$\chi^2$ regions at their $x$ values are ``thin'' in the $y$ direction.   A procedure that excludes such ``thin-contour'' models, in favor of ``thick-contour'' models such as $B$, can be problematic for two reasons.  First of all, it is not clear that low-$\chi^2$ models such as $A$ really should be ignored; it is prudent, at least, to know of the existence of such a model.  Points at the ends of long, narrow regions of degeneracy are ``thin-contour'' points, so marginalization can underestimate the sizes of these degenerate regions.  Such points should be considered in a proper treatment of parameter degeneracies.  

Secondly, the notions of ``thinness'', ``thickness'', and ``distance'' in parameter space depend on the prior constraints chosen.  For a parameter vector $\mathbf{p}$ of $N$ parameters, the {\em{a posteriori}} probability marginalized over all but $m$ of the parameters is given by
\begin{equation}
{\mathcal{P}}(p^0,\dots,p^{m-1}) = \int dp^m \dots dp^{N-1} {\mathcal{L}}({\mathbf{p}}) {\mathcal{P}}_{prior}({\mathbf{p}}).
\label{eqn:marginalize}
\end{equation}
It is customary to reduce this dependence on priors by choosing uniform priors, with ${\mathcal{P}}_{prior}({\mathbf{p}})$ constant on a subset of parameter space and zero elsewhere.  However, the concept of uniform priors is itself parameterization-dependent.  Choosing priors that are uniform in a different set of parameters $\mathbf{q}$ is equivalent to letting ${\mathcal{P}}_{prior}({\mathbf{p}})$ equal the Jacobian determinant of the transformation from $\mathbf{p}$ to $\mathbf{q}$.  

\begin{table}[tb]  
\begin{center}
\begin{tabular}{|c|c|c|}
\hline
contour        &   $\chi^2 - \chi^2_{min}$   &  probability \\
\hline
\hline
$n \sigma$     &   $n^2$                     &  \\ 
\hline 
$1 \sigma$     &   $1$                       &  $39.3 \%$ \\   
\hline
$1.52 \sigma$  &   $2.3$                     &  $68.3 \%$ \\ 
\hline 
$2 \sigma$     &   $4$                       &  $86.5 \%$ \\ 
\hline
$2.49 \sigma$  &   $6.2$                     &  $95.5 \%$ \\ 
\hline
\end{tabular} 
\end{center}
\caption{Comparison between $n \sigma$ $\chi^2$ contours and probability contours for a Gaussian likelihood function of two variables.  \label{tab:sigVsProb}}
\end{table}

On the other hand, marginalization has several advantages of its own.  First of all, one can always come up with a $\chi^2(x,y)$ function which is handled correctly by marginalization, but for which minimization implies artificially tight constraints.  Also, marginalization naturally defines a probability in parameter space.  An analysis based on marginalization is capable of providing probability contours, which may be more useful than $\chi^2$ contours.  If the likelihood function is Gaussian, these two types of contours are related in a simple way, as shown in Table \ref{tab:sigVsProb}.  However, for a non-Gaussian likelihood function, there is no simple way to relate contours of $\chi^2(\mathbf{p})$ to a given probability contour without marginalization.  Meanwhile, it may be the case that a ``natural'' set of parameters exists for describing a theory, such as $h$, $\Omega_m$, and $\Omega_b$ in general relativistic cosmology.  The existence of ``natural'' parameters makes it easier to choose a ``reasonable'' prior probability distribution ${\mathcal{P}}_{prior}$, since one no longer needs to worry about reparameterization changing the form of ${\mathcal{P}}_{prior}$.  Constraints derived from a marginalization over such parameters, assuming weak and uniform priors, can be quite convincing.

We take the point of view that any final constraints on $w_0$ and $w_1$ should be independent of the analysis method used.  If a claim made using one of the two methods does not hold up to scrutiny by the other method, then the issue is too close to call.  For example, we do not believe that model $A$ in Fig. \ref{fig:minVsMarg} is ruled out, even though it is excluded by marginalization over $y$.  In this sense, $\chi^2$ minimization and marginalization are complementary analysis techniques that are useful for handling different types of parameter degeneracies in a non-Gaussian likelihood function.  We have chosen to use $\chi^2$ minimization, partly because we are concerned about degenerate regions such as in \ref{fig:minVsMarg}(top), and partly because it is not clear which parameters should be used to describe dark energy, reionization ($\tau$ or $z_{reion}$), and the power spectrum amplitude ($A$ or $\sigma_8$).

For completeness, we discuss a few computational issues here.   Minimizations were performed using the Amoeba routine of \cite{num_recipes}, although the Powell routine from that reference was found to give very similar results.  The computation was sped up by generating CMB power spectra for $100$ different $n_s$ values at once, and interpolating to find $\chi^2$ for intermediate $n_s$ values.  This gave us information about $\chi^2$ over our entire range of $n_s$ values with just one call to CMBFAST.  (Interpolation-related errors in $\chi^2$ were found to be negligible.)  Also, we minimized separately over the three parameters $A$, $n_s$, and $z_s$, since a variation in one of these parameters did not require the recomputation of CMB power spectra.  This minimization over $A$, $n_s$, and $z_s$ was nested within the minimization over $h$, $\omega_m$, $\omega_b$, and $\tau$.  Using this technique, $\chi^2_{CMB}$ could be minimized at one point in the ($w_0$, $w_1$) plane in about $2-4$ hours, on a single node of the Hydra computer cluster.


\section{Cosmological Probes}
\label{sec:probes}
\subsection{Current data}
\label{subsec:currentChi2}

Since our analyses of current data follow standard procedures, we will discuss them only briefly.

{\em{Supernovae of Type Ia:}}  Supernovae of type Ia (SNe Ia) are standardizable candles \cite{Phillips, Hamuy, RPK94, RPK95}, with magnitude $m(z)$ at red shift $z$ given in, {\em{e.g.}}, \cite{Perl97}, by
\begin{equation}
m = 5\: \log_{10}(D_L) + \mathcal{M} .
\label{eqn:m_vs_DL}
\end{equation}
The dimensionless luminosity distance $D_L(z) = (1+z) H_0 \int_0^z dz'/H(z')$ depends on $\Omega_m$ and the dark energy parameters through $H(z)$, given in (\ref{eqn:H_vs_z}).  Our analysis minimizes $\chi^2$ with respect to the magnitude parameter $\mathcal{M}$, which is dependent on $H_0$ and the SN Ia absolute magnitude.

{\em{Cosmic Microwave Background:}}  For the dynamical models of dark energy considered here, power spectra were computed with a modified version of CMBFAST 4.1 \cite{CMBFAST}, provided by R. Caldwell \cite{Caldwell}.  Our analysis of the data used our own implementation of the CMB $\chi^2$ function described in \cite{Verde}.  Noise parameters and constants describing the WMAP parameterization of the Fisher matrices were taken from the data tables created by \cite{Hinshaw} and \cite{Kogut}, and provided with the WMAP likelihood code. 

{\em{Galaxy Power Spectrum:}}  Our analysis of Sloan Digital Sky Survey (SDSS) data used the galaxy power spectrum \cite{SDSS_pwrspec} and the likelihood code of \cite{SDSS_params}, made publically available at \cite{SDSS_likelihood}.  Given a cosmological model, the corresponding matter power spectrum was computed using CMBFAST.  The normalization of the power spectrum was treated as a nuisance parameter; $\chi^2$ was minimized with respect to it.  An accurate formula for the matter power spectrum in the nonlinear regime, for dynamical dark energy cosmologies, is not currently available.  Furthermore, galaxy biasing in the nonlinear regime is even less well understood.  Therefore, the SDSS data analysis presented here only used measurements for $k \leq 0.15 h/{\textrm{Mpc}}$, as recommended in \cite{SDSS_params}.  


\subsection{SN Ia Simulation}
\label{subsec:snMC}
\subsubsection{Supernova simulation strategies}

Monte Carlo techniques were used to simulate the magnitudes and red shifts of Type Ia Supernovae, from previous, current, and future supernova surveys, that would be available for analysis by the end of the decade.  Rather than mixing real and simulated data, it was decided to simulate all of the surveys from scratch.  In order to conduct a realistic simulation of future supernova surveys, we studied the available literature, including analyses of current data as well as plans for future surveys.  For each SN Ia survey we estimated the expected number of supernovae to be observed and modeled a SN Ia red shift distribution.  The final simulated SN Ia data set included 2050 supernovae ranging in red shift from $0.01$ to $2.0$, as summarized in Table \ref{tab:snList}.  Our simulation was based upon the following assumptions, which we justified by reference to data whenever possible.
\begin{itemize}
     \item $25\%$ of the SNe Ia found by each survey were assumed to be unusable for cosmological analysis.  This is based on the fact that $23\%$ of the SNe Ia in the Tonry/Barris data set \cite{Tonry, Barris} did not survive their cuts on very low red shifts ($z<0.01$) and galactic host extinctions.
     \item Two-thirds of all low red shift SNe (of all types) were assumed to be of type Ia, based on the supernovae reported in \cite{SNfactory}.
     \item Red shift distributions were assumed to be uniform for low red shifts.  Since $m(z)$ at low red shifts is independent of cosmological parameters other than $\mathcal{M}$, the details of the red shift distribution for $z \lesssim 0.1$ should be unimportant for parameter constraints.
     \item Red shift distributions were assumed to be Gaussian ($P(z) \propto \exp \left( -\frac{(z-z_{mean})^2}{2 \sigma_z^2} \right)$) for higher red shift surveys, unless otherwise specified in the literature.
     \item Current data reported in \cite{Tonry, Barris, Knop, Riess04} were simulated by dividing them into three ``surveys:'' 
          \begin{itemize}
               \item[-] low $z$, Tonry/Barris SNe Ia with $z<0.1$;
	       \item[-] mid $z$, SNe Ia from \cite{Tonry, Barris, Knop} with $z>0.1$;
	       \item[-] high $z$, HST GOODS SNe Ia \cite{Riess04}.
          \end{itemize}
          This was done because the current data come from a collection of previous surveys over a wide range of red shifts, and are poorly approximated by uniform or Gaussian distributions.
     \item $40\%$ of the expected Dark Energy Camera \cite{DECamera} data were assumed to be available, bringing the total number of supernovae to just over 2000.  Even under optimistic assumptions about systematic uncertainties, we do not expect an increase in the number of SNe beyond this point to improve constraints on dark energy \cite{Kim04}.
\end{itemize}

\subsubsection{Magnitude uncertainties}
A well-studied SN Ia sample, with accurate spectroscopic measurements of the supernova host galaxy red shifts, can have an intrinsic magnitude uncertainty for each supernova as low as $\sigma_m^{(int)} = 0.15$ \cite{SNLS3, DECamera}.  If less accurate   host red shift information is available, the magnitude uncertainy suffers; \cite{DECamera} estimates an uncertainty of $\sigma_m^{(int)} \sim 0.25$ when only photometric red shift information is available for the host galaxies.  Meanwhile, the average magnitude uncertainty in the Tonry/Barris sample, for SNe Ia in the red shift range $0.1<z<0.8$, is $\sigma_m^{(int)} = 0.3$.  Since the final dataset will be a combination of many supernovae with varying amounts of red shift information, the approach adopted here is to assume a magnitude uncertainty of $\sigma_m^{(int)} = 0.2$ for supernovae in the intermediate red shift range $0.1<z<0.8$. 

Judging from the supernovae tabulated in \cite{Tonry} and \cite{Barris}, the low red shift supernovae tend to have smaller magnitude uncertainties than average.  The mean magnitude uncertainty is $0.18$ for the $z<0.1$ SNe Ia, compared to $0.25$ for the full Tonry/Barris sample.  Thus, for the purposes of this simulation, it is assumed that the magnitude uncertainty for each low red shift ($z<0.1$) supernova is the minimum value, $\sigma_m^{(int)} = 0.15$.  

Meanwhile, supernovae at red shifts greater than about $0.8-1.0$ become increasingly difficult to observe from the ground, and the spectroscopic analyses of such SNe Ia become very time-consuming \cite{SNLS3, Stubbs}.  In the Tonry/Barris sample, the magnitude uncertainty rises from a mean of $0.3$ in the red shift range $0.1<z<0.8$ to $0.35$ in the range $z>0.8$.  Based on this, it was decided to assume an intrinsic magnitude uncertainty of $\sigma_m^{(int)}=0.25$ for supernovae with $z>0.8$.  
\begin{figure}[tb]
\begin{center}
\includegraphics[width=2.3in, angle=270]{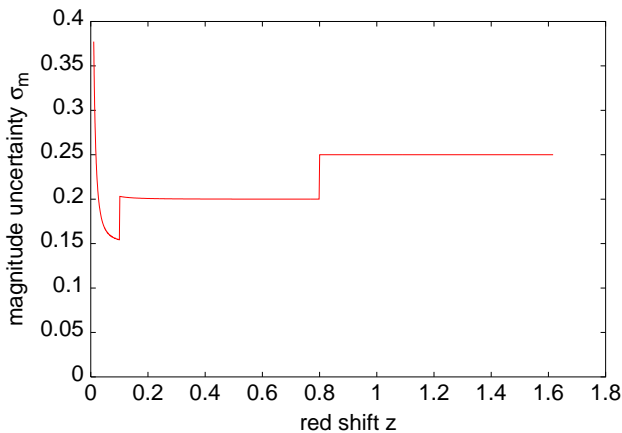}
\caption{Supernova apparent magnitude uncertainty used in the simulation}
\label{fig:sigma_m}
\end{center}
\end{figure}

The correspondence (\ref{eqn:m_vs_DL}) between apparent magnitude and red shift is exact only in the absence of a peculiar velocity.  It is necessary to include an extra uncertainty to account for the nonzero peculiar velocities of actual supernovae.  Following \cite{Tonry}, it was assumed here that the peculiar velocity uncertainty of each supernova was $\sigma_v=500\;km/sec$.  The fractional uncertainty in the luminosity distance due the peculiar velocity uncertainty was taken to equal the fractional uncertainty $\frac{\sigma_v/c}{z}$ in red shift.  The resulting contribution to the magnitude uncertainty is 
\begin{equation}
\sigma_m^{(pec)} = \frac{5}{\ln(10)} \frac{\sigma_v}{c \, z},
\label{eqn:sigma_m_pec}
\end{equation}
where $c$ is the speed of light.  The resulting form of the magnitude uncertainty, after adding $\sigma_m^{(pec)}$ in quadrature to the intrinsic uncertainty, is plotted in Fig. \ref{fig:sigma_m}.  

Finally, following \cite{Kim04}, we assumed a systematic uncertainty that prevents constraints on the apparent magnitude $m(z)$ from becoming arbitrarily tight.  In the absence of such a systematic effect, the magnitude uncertainty in a bin containing $N_{bin}$ supernovae will be $\sigma_m / \sqrt(N_{bin})$, which approaches zero with increasing $N_{bin}$.  To this we add $\delta m = 0.04$ in quadrature, which imposes a floor $\delta m$ on the uncertainty in each bin.  We choose the bin size $\Delta z = 0.1$; this acts as a ``correlation length'' for the systematic.  For a single supernova in a bin containting $N_{bin}$ supernovae, this systematic implies an effective magnitude uncertainty
\begin{equation}
\sigma_m^{(eff)} = \sqrt{ \left(\sigma_m^{(int)}\right)^2 
                          + \left(\sigma_m^{(pec)}\right)^2
                          + N_{bin} \delta m ^2 },
\label{eqn:sigma_m_eff}
\end{equation}
which increases with increasing $N_{bin}$.

\subsubsection{The complete SNe Monte Carlo dataset}

\begin{table}[tb]
\begin{center}
\begin{tabular}{|l||c|c|}
\hline
SN Ia                        & number    & red shift     \\
data set                     & of SNe Ia & distribution \\
\hline
\hline 

Current (low $z$)            & 83        & uniform, \\
\cite{Tonry, Barris}         &           & $0.01<z<0.1$ \\
\hline

Current (mid $z$)            & 117       & Gaussian, \\
\cite{Tonry, Barris, Knop}   &           & $z_{mean}=0.55$, $\sigma_z=0.25$ \\
\hline

Current (high $z$)           & 16        & Gaussian, \\
\cite{Riess04}               &           & $z_{mean}=1.0$, $\sigma_z=0.4$ \\
\hline

Carnegie (low $z$)           & 94        & uniform, \\
\cite{CarnegieSN}            &           & $0.01<z<0.07$\\
\hline

Carnegie (high $z$)          & 90        & Gaussian, \\
\cite{CarnegieSN}            &           & $z_{mean}=0.4$, $\sigma_z=0.2$ \\
\hline

DE Camera                    & 570       & Gaussian, \\
\cite{DECamera}              &           & $z_{mean}=0.55$, $\sigma_z=0.25$\\
\hline

ESSENCE                      & 150       & uniform, \\
\cite{ESSENCE1, ESSENCE2, ESSENCE3} &    & $0.15<z<0.75$\\
\hline

PANS                         & 80        & Gaussian, \\
\cite{PANSabs, PANSwww}      &           & $z_{mean}=1.0$, $\sigma_z=0.4$ \\
\hline

SDSS                         & 100       & uniform, \\
\cite{SDSS}                  &           & $0.05<z<0.15$ \\
\hline

SNfactory                    & 225       & uniform, \\
\cite{SNfactory}             &           & $0.01<z<0.1$ \\
\hline

SNLS                         & 525       & Gaussian, \\
\cite{SNLS1, SNLS2, SNLS3, SNLS4} &      & $z_{mean}=0.6$, $\sigma_z=0.3$ \\
\hline

\hline
Total & 2050 & \\
\hline
\end{tabular}
\caption{Numbers and red shift distributions of simulated type Ia supernovae. \label{tab:snList}}
\end{center}
\end{table}

\begin{figure}[tb]
\begin{center}
\includegraphics[width=2.3in,angle=270]{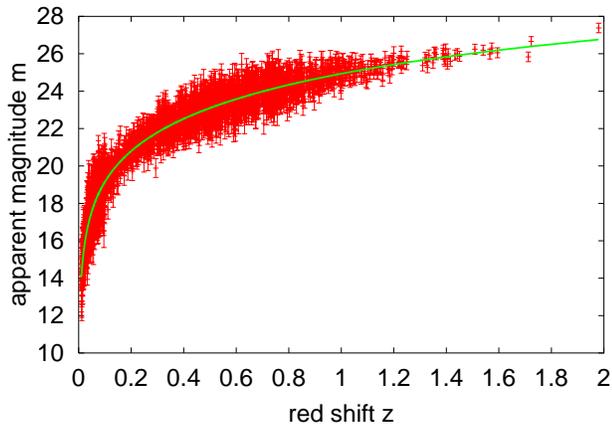}
\caption{The simulated SN Ia dataset.  The points with errorbars are the simulated supernovae, while the solid line is the fiducial model.}
\label{fig:snDataSet}
\end{center}
\end{figure}

Monte Carlo supernova ``data'' points were generated using the red shift distributions and magnitude uncertainties specified above.  For each supernova from each simulated experiment, the appropriate red shift distribution was used for the random generation of a red shift $z$.  (To be specific, the red shift probability distribution $P_i(z)$ for each experiment $i$ was first normalized so that its maximum value was 1.  Then, two random numbers, $r_1$ and $r_2$, were chosen from a uniform distribution between 0 and 1.  If the probability $P_i(z_{max}r_1)$ at red shift $z_{max}r_1$ was greater than $r_2$, then $z_{max}r_1$ was taken to be the red shift of the supernova.  Otherwise, the two numbers were discarded and the process repeated.  Here, $z_{max}$, the maximum allowed supernova red shift, was set to $2.0$ for each experiment.)  Next, the apparent magnitude $m^{(fid)}(z)$ expected for a supernova at red shift $z$ was computed for the fiducial model used here:  $\Omega_m = 0.31$, $w_0=-1$, $w_1=0$.  The magnitude uncertainty $\sigma_m^{(eff)}(z)$ for that supernova was computed as described in the previous subsection.  Finally, the simulated apparent magnitude $m^{(MC)}$ was computed by adding to $m^{(fid)}(z)$ a random number chosen from a Gaussian distribution, with mean zero and standard deviation $\sigma_m^{(eff)}$.  These Monte Carlo simulated magnitudes $m^{(MC)}$, along with their corresponding red shifts and magnitude uncertainties, formed the simulated dataset used here.

Table \ref{tab:snList} lists the numbers and red shift distributions of the supernovae simulated, and Fig. \ref{fig:snDataSet} plots the magnitude-red shift relation of the simulated ``data''.  Note that, since the major supernova surveys planned over the next several years are ground-based, the simulated dataset contains very few supernovae with red shift greater than about $1.2$.  


\subsection{CMB Simulation}
\label{subsec:cmbMC}
\subsubsection{CMB covariance matrix}

In the following discussion, we use the convention of \cite{Bond2000} that ${\mathcal{C}}_l = l(l+1)C_l/(2\pi)$ for any power spectrum $C_l$.  We computed variances in the ${\mathcal{C}}_l$s using the expressions given in \cite{Verde} and \cite{Kogut} for the diagonal components of the covariance matrices.  The noise parameters ${\mathcal{N}}_l^{TT}$ and ${\mathcal{N}}_l^{EE}$ were defined to be $1/8$ of those used by WMAP, in order to simulate the effects of using eight years of WMAP data.  WMAP noise parameters were taken from their data tables, which were described in \cite{Hinshaw} and \cite{Kogut}, and provided along with the WMAP likelihood code described in \cite{Verde}.  The effective noise parameter ${\mathcal{N}}_l^{TT,(eff)}$ as a fraction of the original noise parameter ${\mathcal{N}}_l^{TT}$, and the effective sky fraction $f_{sky}^{(eff)}(l)$, were defined as in \cite{Verde}.  The final expression for the uncertainties in the ${\mathcal{C}}_l^{TT}$s was therefore taken to be
\begin{equation}
\sqrt{Var(TT)} 
= \frac{{\mathcal{C}}_l^{TT,(theory)}+ {\mathcal{N}}_l^{TT,(eff)}}{\sqrt{(l+\frac{1}{2})f_{sky}^{(eff) 2}}} .
\label{eqn:dCTT}
\end{equation}
Note that (\ref{eqn:dCTT}) is a more conservative expression for the variance than that given by \cite{Bond2000}, since there is an ``extra'' factor of $f_{sky}$ in the denominator of $Var(TT)$.   The information ``lost'' by dividing $Var(TT)$ by $f_{sky}$ is actually stored in the off-diagonal components of the covariance matrix (which we ignored), because nearby ${\mathcal{C}}_l$s are anticorrelated \cite{Verde}.  However, we found that this extra $f_{sky}$ had only a negligible effect on our forecast constraints.

We used a similar form for the uncertainty in the $E$-mode  polarization power spectrum  ${\mathcal{C}}_l^{EE}$.  Since WMAP did not include the  ${\mathcal{C}}_l^{EE}$s in the likelihood analysis of their first year data, effective corrections to the noise and sky fraction were not computed.  We used $f_{sky}=0.85$ and ${\mathcal{N}}_l^{EE,(eff)}={\mathcal{N}}_l^{EE}$.
\begin{equation}
\sqrt{Var(EE)}
= \frac{{\mathcal{C}}_l^{EE,(theory)}+ {\mathcal{N}}_l^{EE}}{\sqrt{(l+\frac{1}{2})f_{sky}^{\phantom{sky}2}}} 
\label{eqn:dCEE}
\end{equation}
Finally, uncertainties for the cross-power spectrum ${\mathcal{C}}_l^{TE}$ were taken from Eq. (10) in \cite{Kogut}, using $f_{sky}=0.85$ and $f_{sky}^{TE,eff}=\frac{f_{sky}}{1.14}$ as described in that reference.
\begin{equation}
Var(TE) 
=  \frac{({\mathcal{C}}_l^{TT}+{\mathcal{N}}_l^{TT})({\mathcal{C}}_l^{EE}+{\mathcal{N}}_l^{EE}) + ({\mathcal{C}}_l^{TE})^2}{(2l+1)f_{sky}f_{sky}^{TE,eff}}  
\label{eqn:dCTE}
\end{equation}

Even in the absence of a sky cut, correlations exist among the three power spectra ${\mathcal{C}}_l^{TT}$,  ${\mathcal{C}}_l^{TE}$, and ${\mathcal{C}}_l^{EE}$ for each $l$ \cite{Matias1997}.  As above, we keep the ``extra'' factor of $f_{sky}$ in the denominators of the expressions given in that reference.  For a given multipole $l$, the covariance matrix ${\mathbf{C}}_{XY}$, where $X,Y = TT, \, TE, \, or \, EE$, is given by,
{\setlength\arraycolsep{2pt}
\begin{eqnarray}
\label{eqn:CovXX}
{\mathbf{C}}_{XX}^{(l)} & = & Var(X) \\
\label{eqn:CovTTTE}
{\mathbf{C}}_{TT,TE}^{(l)} & = & \frac{({\mathcal{C}}_l^{TT}+{\mathcal{N}}_l^{TT}){\mathcal{C}}_{l}^{TE}}{(l+\frac{1}{2}) f_{sky}^{\phantom{sky}2}} \\
\label{eqn:CovTTEE}
{\mathbf{C}}_{TT,EE}^{(l)} & = & \frac{({\mathcal{C}}_l^{TE})^2}{(l+\frac{1}{2}) f_{sky}^{\phantom{sky}2}} \\
\label{eqn:CovTEEE}
{\mathbf{C}}_{TE,EE}^{(l)} & = & \frac{({\mathcal{C}}_l^{EE}+{\mathcal{N}}_l^{EE}){\mathcal{C}}_{l}^{TE}}{(l+\frac{1}{2}) f_{sky}^{\phantom{sky}2}} 
\end{eqnarray}}

Large-$l$ correlations are also expected to exist between the CMB power spectra and large scale structure, due to the Sunyaev-Zeldovich effect \cite{SZ1969, SZ1972} and the gravitational lensing of the CMB \cite{Blanchard, Kashlinsky}.  However, these effects are too small to detect today \cite{Huff2004, Uros-CMB_LSS, Chris-CMB_LSS}.  The analysis presented here neglected such correlations.  In particular, we assumed the CMB $\chi^2$ function to be independent of the red shifts $z_s$ of weak lensing sources.


\subsubsection{Simulation}
A set of simulated TT, TE, and EE CMB power spectra was generated by means of a Monte Carlo simulation.  For the purposes of this simulation, it was assumed that the WMAP project \cite{Bennett} would be extended to eight years.  Data from the Planck mission, as described in \cite{Planck1} and \cite{Planck2}, were not simulated for this work; the analysis of Planck data will be complicated by nonlinear effects such as the gravitational lensing of the power spectra.

\begin{figure}[tb]
\begin{center}
\includegraphics[width=2.4in, angle=270]{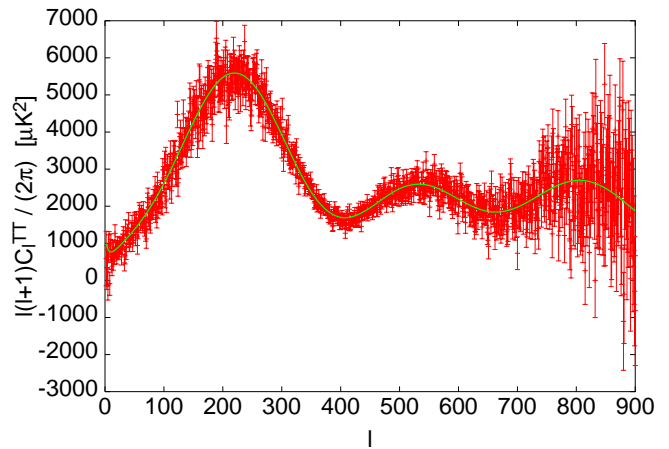}
\caption{WMAP-8 simulated TT power spectrum.  The points with errorbars are the simulated ``data'', while the solid line is the fiducial model.}
\label{fig:CTTmc}
\end{center}
\end{figure}

\begin{figure}[tb]
\begin{center}
\includegraphics[width=2.4in, angle=270]{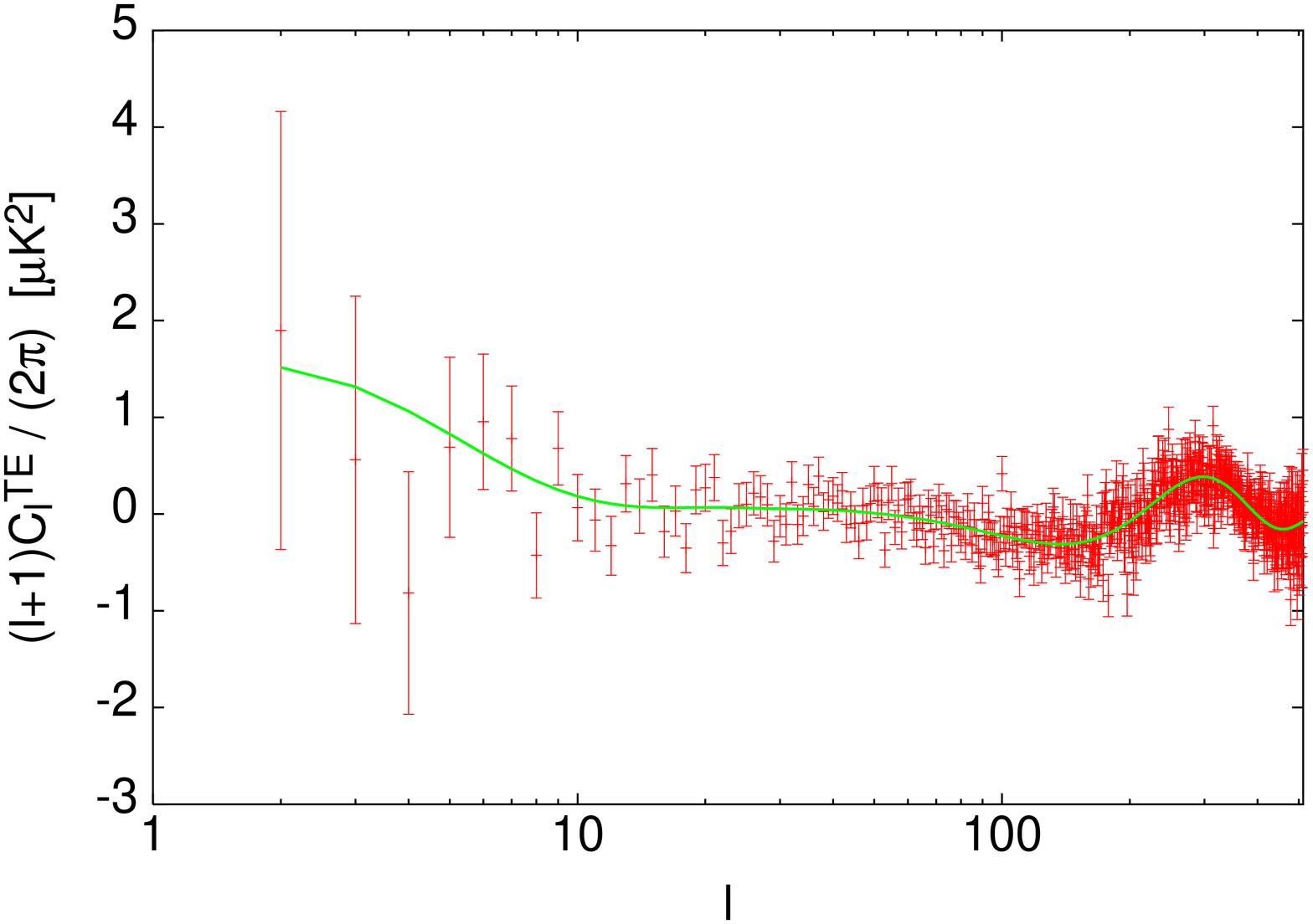}
\caption{WMAP-8 simulated TE power spectrum.  The points with errorbars are the simulated ``data'', while the solid line is the fiducial model.}
\label{fig:CTEmc}
\end{center}
\end{figure}

\begin{figure}[tb]
\begin{center}
\includegraphics[width=2.4in, angle=270]{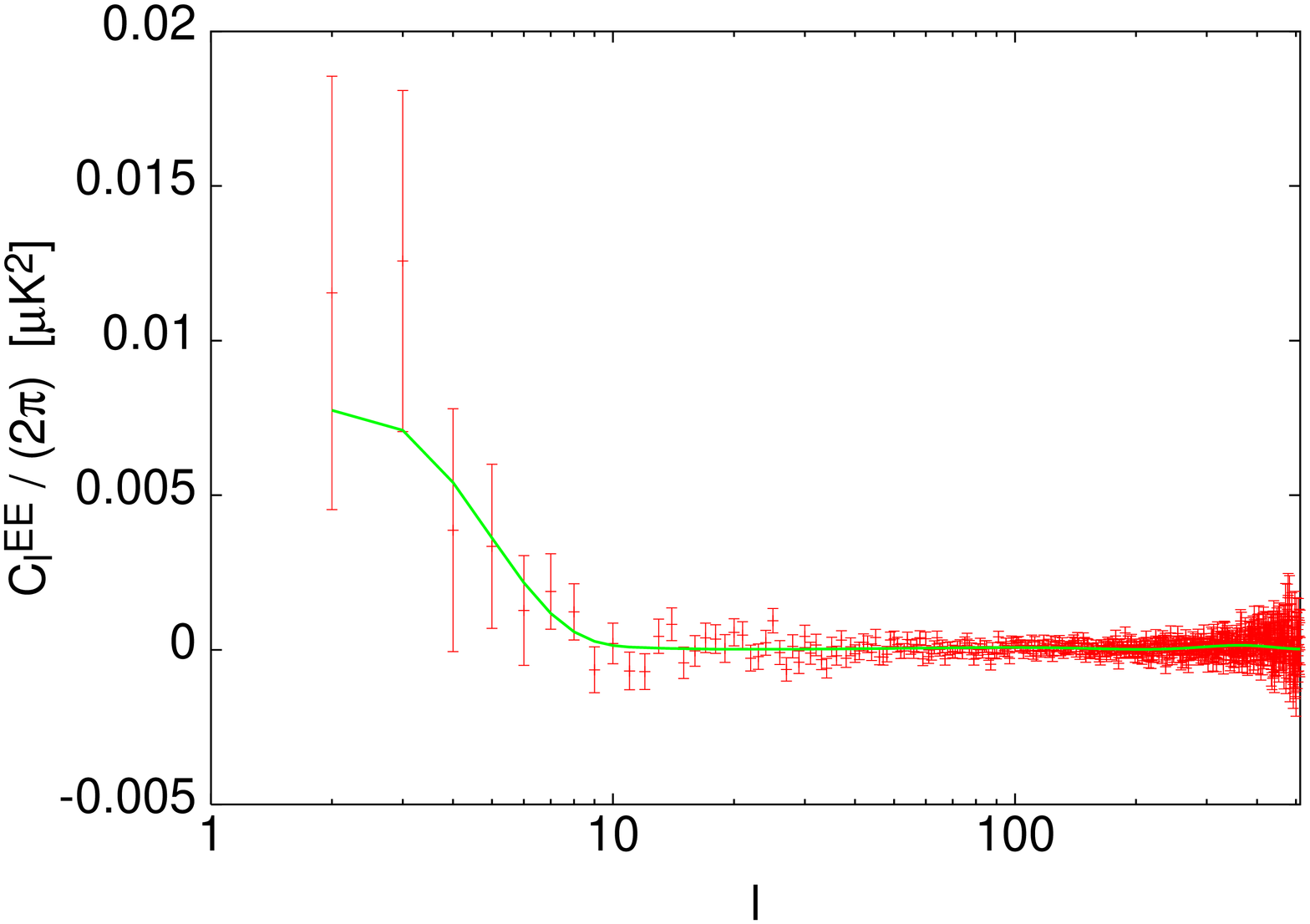}
\caption{WMAP-8 simulated EE power spectrum.  The points with errorbars are the simulated ``data'', while the solid line is the fiducial model.}
\label{fig:CEEmc}
\end{center}
\end{figure}

The WMAP-8 data set described here contained eight years of simulated WMAP data.  We simulated $TT$ data up to a maximum multipole of $l_{max}^{T}=900$, and $TE$ and $EE$ data up to $l_{max}^{E}=512$.  As detailed above, we divided by eight the tabulated WMAP noise parameters from one year of data.  Given the CMBFAST-generated fiducial power spectra ${\mathcal{C}}_l^{TT,(fid)}$, ${\mathcal{C}}_l^{TE,(fid)}$, and ${\mathcal{C}}_l^{EE,(fid)}$, as well as our simulated noise parameters, the power spectrum covariance matrices (\ref{eqn:CovXX}-\ref{eqn:CovTEEE}) were computed.  For each $l$, the inverse covariance matrix ${\mathbf{C}^{(l)}}^{-1}$ was diagonalized,
\begin{equation}
\mathbf{D}^{(l)} = {\mathbf{U}^{(l)}}^{-1} {\mathbf{C}^{(l)}}^{-1} {\mathbf{U}^{(l)}} 
           = \left( \begin{array}{ccc}
                                      \frac{1}{(\delta u_l^1)^2} & 0 & 0 \\
				      0 & \frac{1}{(\delta u_l^2)^2} & 0 \\
				      0 & 0 & \frac{1}{(\delta u_l^3)^2} 
                    \end{array} \right),     
\label{eqn:diagInvCov}
\end{equation}
using the matrix ${\mathbf{U}^{(l)}}$ of eigenvectors of ${\mathbf{C}^{(l)}}^{-1}$.  Next, ${\mathbf{U}^{(l)}}$ was used to rotate the vector of fiducial power spectra into the eigenbasis, 
\begin{equation}
\left( \begin{array}{c} u_l^1 \\ u_l^2 \\ u_l^3 \end{array} \right) = 
{\mathbf{U}^{(l)}} 
\left( \begin{array}{c} {\mathcal{C}}_l^{TT,(fid)} \\ {\mathcal{C}}_l^{TE,(fid)} \\ {\mathcal{C}}_l^{EE,(fid)} \end{array} \right).  
\label{eqn:rot_into_eigenbasis}
\end{equation}
The Monte Carlo simulated power spectra were generated in the eigenbasis, using random numbers $r_1$, $r_2$, and $r_3$ chosen from a Gaussian distribution with zero mean and unit variance:
{\vbox{\setlength\arraycolsep{2pt}
\begin{eqnarray}
u_l^{1,(MC)} & = & u^1_l + r_1 \, \delta u_l^1 \\
u_l^{2,(MC)} & = & u^2_l + r_2 \, \delta u_l^2 \\
u_l^{3,(MC)} & = & u^3_l + r_3 \, \delta u_l^3 .
\end{eqnarray}}}
Finally, the Monte Carlo power spectra were rotated back into the standard basis, 
\begin{equation}
\left( \begin{array}{c} {\mathcal{C}}_l^{TT,(MC)} \\ {\mathcal{C}}_l^{TE,(MC)} \\ {\mathcal{C}}_l^{EE,(MC)} \end{array} \right) = 
{\mathbf{U}^{(l)}}^{-1} 
\left( \begin{array}{c} u_l^{1,(MC)} \\ u_l^{2,(MC)} \\ u_l^{3,(MC)} \end{array} \right) .
\label{eqn:rot_into_std_basis}
\end{equation}
These final simulated power spectra, ${\mathcal{C}}_l^{TT,(MC)}$, ${\mathcal{C}}_l^{TE,(MC)}$, and ${\mathcal{C}}_l^{EE,(MC)}$, were the ones used in our analysis.  Note that this simulation method ignores non-Gaussianities in the distributions of observed $C_l$ values about the underlying model; see \cite{Bond2000} for a discussion of such effects.  The simulated TT, TE, and EE power spectra are shown in Figs. \ref{fig:CTTmc}, \ref{fig:CTEmc}, and \ref{fig:CEEmc}, respectively.

\subsection{Weak Lensing Simulation}
\label{subsec:wlMC}
\subsubsection{Convergence power spectrum}

In the Limber approximation, the convergence power spectrum
is given by
\cite{1992ApJ...388..272K,1997ApJ...484..560J,1998ApJ...498...26K}:
\begin{equation}
P^{\kappa}_l =
\frac{9}{4} H_0^4\Omega_m^2\int^{\chi_H}_{0}
\frac{g^2(\chi)}{a^2(\chi)}P_{3D}
\left(\frac{l}{\sin_{K}(\chi)},\chi\right) d\chi,
\end{equation}
where $P_{3D}$ is the $3D$ nonlinear power spectrum of the matter
density fluctuation, $\delta$; 
 $a(\chi)$ is the scale factor; and
$\sin_{K}\chi=K^{-1/2}\sin(K^{1/2}\chi)$ is
the comoving angular diameter distance to $\chi$ 
(for the spatially flat universe used in this analysis,
this reduces to $\chi$).
The weighting function $g(\chi)$ is the source-averaged distance ratio
given by
\begin{equation}
\label{eq:weighting}
g(\chi) = \int_\chi^{\chi_H} n(\chi') {\sin_K(\chi'-\chi)\over
\sin_K(\chi')} d\chi',
\end{equation}
where $n(\chi(z))$ is the source red shift distribution normalized by
$\int dz\; n(z)=1$.
We assume that all the sources are at a characteristic red shift $z_s$,
so that (\ref{eq:weighting}) reduces to 
\begin{equation}
\label{eq:weighting2}
g(\chi) =  {\sin_K(\chi_s-\chi)\over
\sin_K(\chi_s)}.
\end{equation}
For weak lensing calculations, we use the standard BBKS transfer function \cite{BBKS}, and the analytic approximation 
of Ref. \cite{ma1999} for the growth factor.
We use the mapping procedure HALOFIT \cite{smith2003} to calculate
the non-linear power spectrum. 


\subsubsection{The Weak Lensing $\chi^2$ function}

For the weak lensing spectrum, the uncertainty is
given by: \cite{1992ApJ...388..272K,1998ApJ...498...26K}
\begin{equation}
\label{eq:conv_error}
\delta P^{\kappa}_l=
\sqrt{\frac{2}{(2 l +1)f_{sky}}}\left (
P^{\kappa}_l + {\left< \gamma_{int}{}^2 \right>\over \bar n} \right ) \,,
\label{delta_kappa}
\end{equation}
\noindent where $f_{sky} = \Theta^2 \pi/129600 $ is the fraction of the
sky covered by a survey of dimension $\Theta$ in degrees, and 
$\left<\gamma_{int}^2\right>^{1/2} \approx 0.4$ is the intrinsic
ellipticity of galaxies. We consider a reference survey 
with $f_{sky}=0.7$, in the same range as the Pan-STARRS survey
\cite{Pan-STARRS}.
We used an average galaxy number density of
$\bar n \approx 6.6 \times 10^{8} \sr^{-1}$.
$\chi^2$ was computed between multipoles $l_{min}$ and $l_{max}$.  For $l_{min}$, we took the fundamental mode approximation:
\begin{equation}
l_{\rm min} \approx \frac{360\rm ~deg}{\Theta} = \sqrt{\pi\over f_{sky} },
\label{eq:lmin}
\end{equation}
i.e. we considered only lensing modes for which at least one wavelength can fit inside
the survey area. We used $l_{max}=3000$, since on smaller scales the nonlinear approximation 
HALOFIT to the nonlinear power spectrum may not be valid.


\subsubsection{Simulation}

The procedure described above was used to generate fiducial 
convergence power spectra $P_l^{\kappa,(fid)}$.  
We used the cosmological parameter fiducial values 
of Sec. \ref{subsec:DEModel}.
We then calculated the uncertainties on the convergence power spectra 
using Eq. (\ref{eq:conv_error}).
The Monte-Carlo simulated convergence power spectra are 
generated using
\begin{equation}
P_l^{\kappa,MC}=P_l^{\kappa,(fid)}+r\;\delta P_l^{\kappa}
\end{equation}
where $r$ is  randomly choosen from a Gaussian distribution of standard
deviation 1 \cite{num_recipes}. 
\begin{figure}[tb]
\begin{center}
\includegraphics[width=2.3in, angle=270]{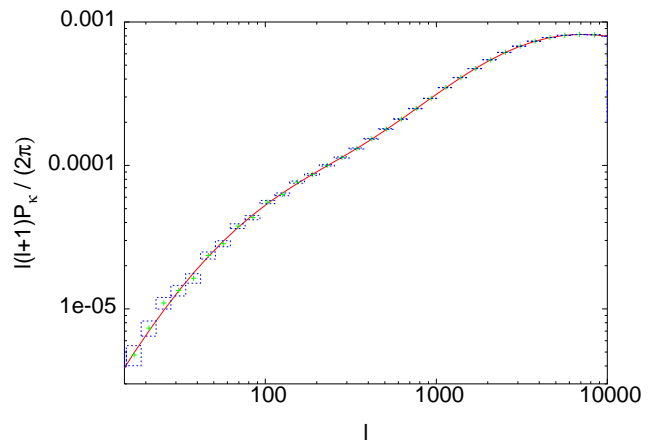}
\caption{Monte Carlo simulated convergence power spectrum.  
The points with error boxes are the simulated ``data''
for a survey with 
$f_{sky}=0.7$, $\bar n \approx 6.6 \times 10^{8} \sr^{-1}$  and 
$\left<\gamma_{int}^2\right>^{1/2} \approx 0.4$,
while the solid line is the fiducial model. \label{fig:lensingmc}}
\end{center}
\end{figure}
The simulated convergence power spectrum for the reference survey and 
the fiducial model  are shown in Fig. \ref{fig:lensingmc}.


\section{Results and discussion}
\label{sec:discussion}
\subsection{Goals}
\label{subsec:goals}
As we have said, the most important question in the study of dark energy today is whether the dark energy is or is not a cosmological constant.  If dark energy is shown not to be a cosmological constant, the next questions that arise are again qualitative: Is the equation of state greater or less than $-1$?  Does it vary with time?  The actual values of $w(0)$ and $w'(0)$ will become important in the more distant future, when theoretical models for the form of $w(z)$ are available.  We look for constraints on the equation of state parameters, not because we believe that any one of the parameterizations (\ref{eqn:EOSsimple}, \ref{eqn:EOSlinder}, \ref{eqn:EOSuros}, \ref{eqn:EOSparam}) is a theoretically favored model for $w(z)$, but because we wish to address the above qualitative questions about dark energy.  Thus, we emphasize that our goal is to answer these qualitative questions in a careful, parameterization-independent way, rather than to find specific values for $w(0)$ and $w'(0)$ within a particular parameterization.


\subsection{Analysis of Current Data}
\label{subsec:current}
\begin{figure}[tb]
\includegraphics[width=3.3in]{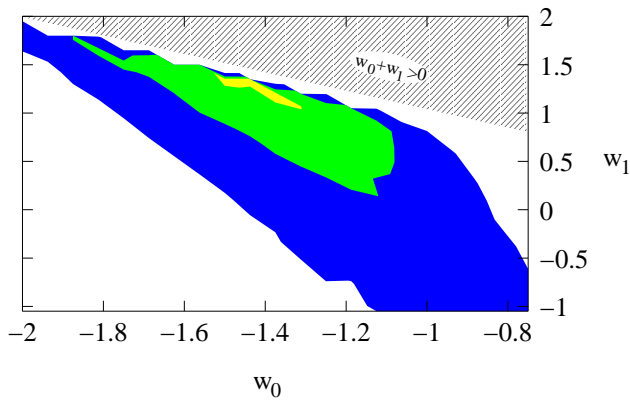}
\caption{Current constraints on dynamical dark energy parameterized by (\ref{eqn:EOSparam}), using first-year WMAP data, the SNe Ia ``gold set'' from \cite{Riess04}, and the current SDSS data.  The $1\sigma$, $2\sigma$, and $3\sigma$ contours are shown.  The region of parameter space excluded by the $w_0+w_1 \leq 0$ prior is filled with upwards-sloping stripes.}
\label{fig:current}
\end{figure}

Keeping these goals in mind, we used $\chi^2$ minimization to analyze current data from the CMB power spectrum \cite{Bennett}, the SN Ia ``gold set'' \cite{Riess04}, and the galaxy power spectrum \cite{SDSS_pwrspec}.   We began by using (\ref{eqn:EOSparam}) to parameterize the dark energy equation of state.  The contours obtained are plotted in Fig. \ref{fig:current}, with grid spacings $\Delta_{w_0}=0.06$ and $\Delta_{w_1}=0.15$.  The resulting $1 \sigma$ and $2 \sigma$ constraints on the dark energy parameters are $w_0=-1.38^{+0.08}_{-0.20}{}^{+0.30}_{-0.55}$ and $w_1=1.2^{+0.40}_{-0.16}{}^{+0.64}_{-1.06}$, with a best-fit $\chi^2$ value of $1611$.  We consider our $2 \sigma$ contours to be more reliable than our $1 \sigma$ contours, since the latter occupy an area in Fig. \ref{fig:current} only a few times the area of a grid square.

$\Lambda$CDM models are not conclusively ruled out by our results.  Compared to the best-fit model in Fig. \ref{fig:current}, the $\Lambda$CDM model has a $\chi^2$ value that is higher by $5$.  However, the cosmological constant is a ``simpler'' model of dark energy than the $w(z)$ parameterizations considered here, in the sense that the $\Lambda$CDM model has two fewer variable parameters than (\ref{eqn:EOSsimple}, \ref{eqn:EOSlinder}, \ref{eqn:EOSparam}).  Thus the $\Lambda$CDM model has $\chi^2/d.o.f. = 1616/1514 = 1.0676$, while the best-fit model from Fig. \ref{fig:current} has $\chi^2/d.o.f. = 1611/1512 = 1.0656$.  In the approximation that uncertainties in the data points are Gaussian, these correspond to probabilities of $P_{\Lambda CDM}=0.0336$ and $P_{best-fit}=0.0378$.  For comparison, a point on the edge of the $2 \sigma$ contour in Fig. \ref{fig:current} has $\chi^2/d.o.f. = 1615/1512 = 1.0682$ and probability $0.0324$.  In this sense, the $\Lambda$CDM model is more probable than some points within the $2 \sigma$ contour.  Although the data favor $w_0<-1$ and $w_1>0$, the cosmological constant is not decisively ruled out as a model of dark energy.

\begin{figure}[tb]
\begin{center}
\includegraphics[width=3.3in]{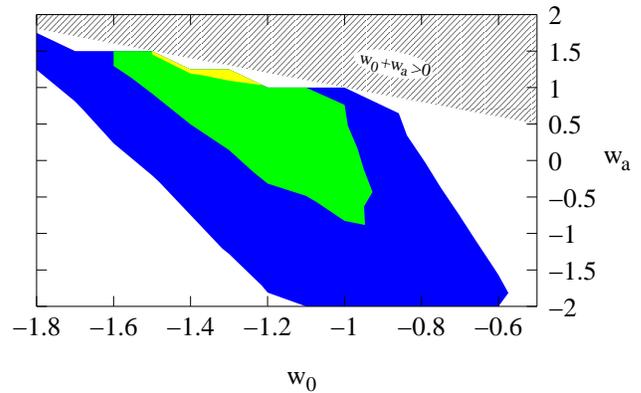}
\caption{Same as Fig. \ref{fig:current}, except that the dark energy parameterization (\ref{eqn:EOSlinder}) has been used.  The region of parameter space excluded by the $w_0+w_a \leq 0$ prior is filled with upwards-sloping stripes.}
\label{fig:linderParam}
\end{center}
\end{figure}

Moreover, we have not considered systematic uncertainties in the supernova data, which can degrade constraints on the equation of state.  As an optimistic estimate of the effects that such a systematic will have on dark energy constraints, we assumed an uncorrelated systematic uncertainty $\delta m = 0.04$, in red shift bins $\Delta z = 0.1$, and reanalyzed the data.  The added uncertainty caused the $2 \sigma$ contour to broaden to include the cosmological constant, while the best-fit model remained the same.  

It is troubling that the best-fit model in Fig. \ref{fig:current} is so close to the boundary $w_0+w_1=0$ of the dark energy parameter space.  This is despite the fact that (\ref{eqn:EOSparam}), out of the three 2-parameter equations of state illustrated in Fig. \ref{fig:wParams}, covers the greatest range of parameters in the ($z$, $w(z)$) plane.  Parametrization (\ref{eqn:EOSlinder}), widely used in the literature,  is more restrictive than (\ref{eqn:EOSparam}), and therefore should have more problems with this boundary.  We repeated our analysis using (\ref{eqn:EOSlinder}), obtaining the $\chi^2$ contours shown in Fig. \ref{fig:linderParam} with grid spacings $\Delta_{w_0}=0.1$ and $\Delta_{w_a}=0.25$.  The resulting $2 \sigma$ bounds are $w_0=-1.3^{+0.39}_{-0.34}$ and $w_a=1.25^{+0.40}_{-2.17}$, with a best-fit $\chi^2$ value of $1613$.  Thus we see that the allowed region shifts significantly when we switch to parameterization (\ref{eqn:EOSlinder}), with $w'(0)<0$ models falling within the $2 \sigma$ contours.  Also, the minimum $\chi^2$ value goes up by $1.8$.  

\begin{figure}[tb]
\begin{center}
\includegraphics[width=3.3in]{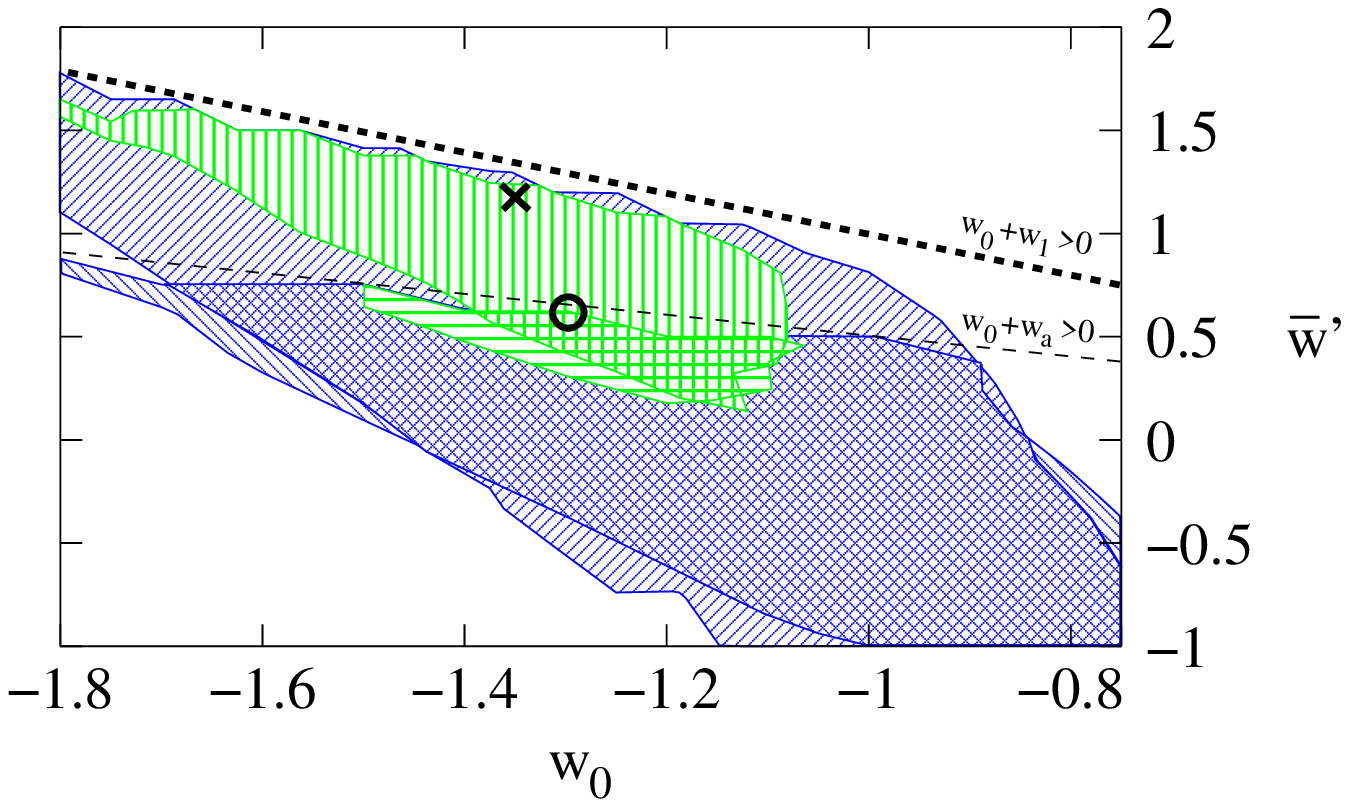}
\caption{Comparison of the two dark energy parameterizations (\ref{eqn:EOSlinder}, \ref{eqn:EOSparam}).  Contours corresponding to (\ref{eqn:EOSparam}) are filled in with vertical stripes ($\chi^2=1615$ contour) or upwards-sloping stripes ($\chi^2=1620$ contour), and the best-fit model is marked by an ``X''.  Contours corresponding to (\ref{eqn:EOSlinder}) are filled in with horizontal stripes ($\chi^2=1615$ contour) or downwards-sloping stripes ($\chi^2=1620$ contour), and the best-fit model is marked by an ``O''.  Note that the $\chi^2=1615$ and $\chi^2=1620$ contours are the $2 \sigma$ and $3 \sigma$ contours, respectively, of parameterization (\ref{eqn:EOSparam}).  The thick and thin dashed lines correspond to the lines $w_0+w_1=0$ and $w_0+w_a=0$, respectively.}
\label{fig:paramComp}
\end{center}
\end{figure} 

These shifts are better understood when the $\chi^2$ functions corresponding to the two parameterizations are plotted together in the ($w_0$, ${\bar{w}}'$) plane (with ${\bar{w}}'\equiv w(1)-w(0)$), as in Fig. \ref{fig:paramComp}.  Note, first of all, that the two sets of $\chi^2$ contours would be nearly identical if the region of parameter space between the dotted lines were removed.  This region corresponds to the slice of the ($z$, $w(z)$) plane between the dashed and solid lines in Fig. \ref{fig:wParams}, which is allowed by parameterization (\ref{eqn:EOSparam}) but not by (\ref{eqn:EOSlinder}).  Also note that the best-fitting point ($w_0=-1.38$, ${\bar{w}}'=1.2$) of parameterization (\ref{eqn:EOSparam}), marked by an ``X'' on the plot, is outside the range allowed by parameterization (\ref{eqn:EOSlinder}).  Since the minimum $\chi^2$ found using (\ref{eqn:EOSlinder}) is higher, the $2 \sigma$ contour is at a higher $\chi^2$, meaning that it now encloses some points with $w_a<0$.  Thus, the difference between the constraints found using (\ref{eqn:EOSparam}) and (\ref{eqn:EOSlinder}) is directly related to the parameterization, and particularly, to the high red shift $w(z)$ constraint discussed in Sec. \ref{subsec:DEModel}.  

To summarize, switching parameterizations from (\ref{eqn:EOSparam}) to (\ref{eqn:EOSlinder}) moves the boundary downwards, pushing the contours in the direction of decreasing ${\bar{w}}'$.  Since the contours in Fig. \ref{fig:current} are still near the boundary, a parameterization that allows larger values of ${\bar{w}}'$ may allow the contours to shift even more in that direction.  It is possible that such a parameterization would favor $w_0<-1$ and ${\bar{w}}'>0$ to an even larger degree, leaving the $\Lambda$CDM model less favored. 

\begin{table*}[tb]   
\begin{center}
\begin{tabular}{|p{0.3in}||p{1.2in}|p{1.5in}|p{1.7in}|p{1.7in}|} 
\hline
Ref.                   & probes included & parameters varied & priors        & dark energy constraints \\
\hline  
\hline 

\cite{Riess04}         & \mbox{\em{SN Ia}}  & $w_0$, $w'$, $\Omega_m$& \mbox{$\Omega_m=0.27\pm0.04$}       
                                                                             & \mbox{$w_0=-1.31^{+0.22}_{-0.28}$,} \mbox{$w'=1.48^{+0.81}_{-0.90}$} \\
\hline

\cite{UrosSdssLyaBias} & {\em{CMB}}, \mbox{\em{SN Ia}}, {\em{$P(k)$}}, {\em{bias}}, {\em{Ly-$\alpha$}}           
                                         & ``vanilla'', $w_0$, $w_a$& \mbox{$\tau<0.3$}       
                                                                             & \mbox{$w_0=-0.981^{+0.193}_{-0.193}{}^{+0.384}_{-0.373}{}^{+0.568}_{-0.521}$,}  \newline  \mbox{$w_a=-0.05^{+0.65}_{-0.83}{}^{+1.13}_{-1.92}{}^{+1.38}_{-2.88}$} \\
\hline

\cite{Hannestad}       & {\em{CMB}}, \mbox{\em{SN Ia}}, {\em{$P(k)$}}          
                                         & ``vanilla'',  \newline 4 $w(z)$ parameters & \mbox{$h=0.72 \pm 0.08$,}  \mbox{$0<\Omega_m<1$,} \mbox{$0.014<\omega_b<0.04$,} \mbox{$0<\tau<1$,} \mbox{$0.6<n_s<1.4$}     
                                                                             & \mbox{$w_0=-1.43^{+0.16}_{-0.38}$,}  \newline \mbox{$\left. \frac{dw}{dz} \right|_{z=0}=1.0^{+1.0}_{-0.8}$} \\

\hline 

\cite{WangTegmark}     & \mbox{\em{SN Ia}}  & $w_0$, $w'$, $\Omega_m$& \mbox{$d_{AC}^{(dec)} \omega_m^{1/2} / c =1.716\pm 0.062$,}   \newline \mbox{$\left.\frac{d \ln D}{d \ln a}\right|_{z=0.15}=0.51\pm0.11$}         
                                                                             & \mbox{$-1.05<w_0<-0.29$,}   \newline \mbox{$-1.89<w'<0.05$} \footnotemark[1] \footnotetext[1]{Bounds taken from plot of $68 \%$ probability contour}\\ 

\hline

\cite{WangTegmark}     & \mbox{\em{SN Ia}}  & $w_0$, $w'$, $\Omega_m$& \mbox{$\Omega_m=0.27\pm0.04$}          
                                                                             & \mbox{$-1.39<w_0<-0.25$,}   \newline \mbox{$-2.61<w'<1.49$} \footnotemark[1]\\

\hline

\cite{Alam}            & \mbox{\em{SN Ia}}  & $\Omega_m$, 3 $w(z)$ parameters &  \mbox{$d_{AC}^{(dec)} \omega_m^{1/2} / c =1.710\pm 0.137$,} \mbox{$\omega_b=0.024$,} \mbox{$\omega_m=0.14 \pm 0.02$}           
                                                                             & \mbox{$\left. \bar{w} \right|_{0<z<0.414}=-1.287^{+0.016}_{-0.056}$,} \mbox{$\left. \bar{w} \right|_{0.414<z<1}=-0.229^{+0.070}_{-0.117}$,} \mbox{$\left. \bar{w} \right|_{1<z<1.755}=0.142^{+0.051}_{-0.033}$} \\

\hline

\cite{Choudhury}       & \mbox{\em{SN Ia}}  & $w_0$, $w_a$& \mbox{$\Omega_m = 0.27$}    & \mbox{$w_0 = -1.38 \pm 0.21$,}   \newline \mbox{$w_a = 2.78 \pm 1.32$} \\

\hline

\cite{Choudhury}       & \mbox{\em{SN Ia}}  & $w_0$, $w_a$& \mbox{$\Omega_m = 0.31$}    & \mbox{$w_0 = -1.35 \pm 0.24$,}  \newline  \mbox{$w_a = 2.32 \pm 1.56$} \\

\hline

\cite{Choudhury}       & \mbox{\em{SN Ia}}  & $w_0$, $w_a$& \mbox{$\Omega_m = 0.35$}    & \mbox{$w_0 = -1.3 \pm 0.29$,}  \newline  \mbox{$w_a = 1.64 \pm 1.95$} \\

\hline

\cite{Rapetti}         & {\em{CMB}}, \mbox{\em{SN Ia}}, \mbox{\em{X-ray}}          
                                         & ``vanilla'', bias, $w_0$, $w_a$& \mbox{bias $=0.824\pm 0.089$,}   \newline \mbox{$\omega_b=0.0214 \pm 0.0020$,}   \newline \mbox{$h=0.72\pm0.08$}
                                                                             & \mbox{$w_0 = -1.16 ^{+0.22}_{-0.19}$,}   \newline  \mbox{$w_0+w_a =-0.05 ^{+0.09}_{-1.17}$} \\

\hline

\cite{Rapetti}         & {\em{CMB}}, \mbox{\em{SN Ia}}, \mbox{\em{X-ray}}             
		                         & ``vanilla'', bias, $\Omega_K$, $w_0$, $w_a$& \mbox{bias $=0.824 \pm 0.089$,}  \newline  \mbox{$\omega_b=0.0214\pm0.0020$,}  \newline  \mbox{$h=0.72\pm0.08$}              
                                                                             & \mbox{$w_0 = -1.14^{+0.31}_{-0.21}$,}  \newline  \mbox{$w_0+w_a = -0.09 ^{+0.12}_{-2.16}$} \\

\hline

\cite{Rapetti}         & {\em{CMB}},  \mbox{\em{SN Ia}}, \mbox{\em{X-ray}}           
                                         & ``vanilla'', bias, \newline  3 $w(z)$ parameters& \mbox{bias $=0.824 \pm 0.089$,}   \newline \mbox{$\omega_b=0.0214\pm0.0020$,}   \newline \mbox{$h=0.72\pm0.08$}        
                                                                             & \mbox{$w_0 = -1.23^{+0.34}_{-0.46}$,}   \newline \mbox{$\lim_{z\rightarrow \infty} w(z) = -0.12 ^{+0.11}_{-0.76}$} \\

\hline

AU, MI, PJS            & {\em{CMB}}, \mbox{\em{SN Ia}}, {\em{$P(k)$}}          
                                         & ``vanilla'', $w_0$, $w_1$& \mbox{for priors see Table \ref{tab:priors}}          
                                                                             & \mbox{$w_0=-1.38^{+0.30}_{-0.55}$,}  \newline  \mbox{$w_1=1.20^{+0.64}_{-1.06}$} \newline \mbox{($2 \sigma$ constraints)} \\

\hline

AU, MI, PJS            & {\em{CMB}}, {\em{$P(k)$}}, \mbox{\em{SN Ia}} (syst. $\delta m = 0.04$)         
                                         & ``vanilla'', $w_0$, $w_1$& \mbox{for priors see Table \ref{tab:priors}}          
                                                                             & \mbox{$w_0=-1.36^{+0.52}_{-0.56}$,}  \newline  \mbox{$w_1=1.20^{+0.60}_{-1.88}$} \newline \mbox{($2 \sigma$ constraints)} \\

\hline

AU, MI, PJS            & {\em{CMB}}, \mbox{\em{SN Ia}}, {\em{$P(k)$}}            
                                         & ``vanilla'', $w_0$, $w_a$& \mbox{for priors see Table \ref{tab:priors}}       
                                                                             & \mbox{$w_0=-1.3^{+0.39}_{-0.34}$,}  \newline  \mbox{$w_a=1.25^{+0.40}_{-2.17}$} \newline \mbox{($2 \sigma$ constraints)} \\
\hline

\end{tabular} 
\end{center}
\caption{Dynamical dark energy constraints from current data. The ``vanilla'' parameter space is spanned by the six parameters ($h$, $\omega_m$, $\omega_b$, $\tau$, $A$, $n_s$). The cosmological probes used are: \mbox{{\em{bias}} = SDSS galaxy  bias,} \mbox{{\em{CMB}} = WMAP CMB power spectra,} \mbox{{\em{Ly-$\alpha$}} = SDSS Lyman-$\alpha$,} \mbox{{\em{$P(k)$}} = SDSS galaxy power spectrum,} \mbox{{\em{SN Ia}} = SN Ia ``gold set'',} \mbox{{\em{X-ray}} = Chandra X-ray clusters}. \label{tab:current}}
\end{table*}  

Comparison with the published literature shows that our results are consistent with others obtained using various combinations of cosmological probes.  Table \ref{tab:current} lists several recent analyses, along with the parameters and $w(z)$ parameterizations used (parameters $w'$, $w_a$, and $w_1$ imply parameterizations (\ref{eqn:EOSsimple}), (\ref{eqn:EOSlinder}), and (\ref{eqn:EOSparam}), respectively), the priors, and the resulting $w(z)$ constraints.  When $w(z)$ constraints using the same equation of state parameterizations are compared, our results are consistent with those in Table \ref{tab:current} at the $2 \sigma$ level.  In addition, the results of \cite{Riess04, WangTegmark}, obtained using parameterization (\ref{eqn:EOSsimple}) and SN Ia data, are consistent with our constraints using either of the parameterizations (\ref{eqn:EOSlinder}, \ref{eqn:EOSparam}).  It is reassuring that our $\chi^2$ minimization procedure, which handles parameter degeneracies differently than marginalization, obtains consistent results.

\begin{table}[tb]   
\begin{center}
\begin{tabular}{|c||c|c|c|}
\hline
$z$      &  Ref.                  &  Parametrization      & $w(z)$ constraint \\
\hline
\hline
$0.3$    & \cite{UrosSdssLyaBias} & (\ref{eqn:EOSlinder}) & $-1.011^{+0.176}_{-0.215}$ ($95.5\%$) \\ \cline{2-4} 
         & \cite{UrosSdssLyaBias} & (\ref{eqn:EOSuros})   & $-0.981^{+0.205}_{-0.249}$ ($95.5\%$) \\ \cline{2-4}
         & AU, MI, PJS            & (\ref{eqn:EOSparam})  & $-1.02^{+0.18}_{-0.33}$ ($2\sigma$) \\ \cline{2-4}
         & AU, MI, PJS            & (\ref{eqn:EOSlinder}) & $-1.01^{+0.19}_{-0.29}$ ($2\sigma$) \\ \cline{2-4}
\hline
$1$      & \cite{UrosSdssLyaBias} & (\ref{eqn:EOSlinder}) & $-1.00^{+0.27}_{-0.66}$ ($95.5\%$) \\ \cline{2-4} 
         & \cite{UrosSdssLyaBias} & (\ref{eqn:EOSuros})   & $-1.03^{+0.39}_{-0.58}$ ($95.5\%$) \\ \cline{2-4}
         & AU, MI, PJS            & (\ref{eqn:EOSparam})  & $-0.18^{+0.12}_{-0.80}$ ($2\sigma$) \\ \cline{2-4}
         & AU, MI, PJS            & (\ref{eqn:EOSlinder}) & $-0.68^{+0.08}_{-0.74}$ ($2\sigma$) \\ \cline{2-4}
\hline 
$\infty$ & \cite{Rapetti}         & (\ref{eqn:EOSlinder}) & $-0.05 ^{+0.09}_{-1.17}$  \\ \cline{2-4}
         & \cite{Rapetti}         & 3 $w(z)$ params.      & $-0.12 ^{+0.11}_{-0.76}$  \\ \cline{2-4}
         & AU, MI, PJS            & (\ref{eqn:EOSparam})  & $-0.18^{+0.12}_{-0.80}$ ($2\sigma$) \\ \cline{2-4}
         & AU, MI, PJS            & (\ref{eqn:EOSlinder}) & $-0.05^{+0.05}_{-1.78}$ ($2\sigma$) \\ \cline{2-4}
\hline
\end{tabular} 
\end{center}
\caption{ Constraints on $w(z)$ at several redshifts. \label{tab:wz_constraints}}
\end{table}  

Several papers list constraints on $w(z)$ at specific values of $z$, as shown in Table \ref{tab:wz_constraints}.  It is evident from the table that $w(0.3)$ is well-constrained and parameterization-independent.  Our results for $w(0.3)$ are in excellent agreement with those of \cite{UrosSdssLyaBias}, even when their three-parameter equation of state (\ref{eqn:EOSuros}) is used.  Meanwhile, the equation of state at $z=1$ is less well constrained, though once again our results agree with those of \cite{UrosSdssLyaBias} at the $2 \sigma$ level.  Parametrization (\ref{eqn:EOSparam}) evidently prefers higher values of $w(1)$ than does (\ref{eqn:EOSlinder}).  One would expect the addition of the third parameter in (\ref{eqn:EOSuros}) to bridge the gap between constraints using (\ref{eqn:EOSlinder}) and (\ref{eqn:EOSparam}).  However, our best-fit models using (\ref{eqn:EOSparam}) have $w(0) \approx -1.4$, $w(0.3) \approx -1$, and $w(1) \approx -0.2$.  Even (\ref{eqn:EOSuros}) cannot reproduce all three of these features if one also imposes the CMB constraint $w(z) \leq 0$ at $z \gg 1$, which implies $w_0+w_a+w_b \leq 0$.  That is, setting $w(0)=-1.38$ and $w(1)=-0.18$ from our best-fit model, and imposing the requirement that $\lim_{z \rightarrow \infty} w(z) \leq 0$, implies that $w(0.3) \geq -0.70 $ in parameterization (\ref{eqn:EOSuros}), which is inconsistent with constraints on $w(0.3)$ discussed above.  Since our best $w$ models are qualitatively different from those allowed using the parameterizations (\ref{eqn:EOSlinder}) and (\ref{eqn:EOSuros}), there is still no discrepancy between our results and those of \cite{UrosSdssLyaBias} for $w(1.0)$  Finally, recall that the two-parameter equations of state (\ref{eqn:EOSsimple}, \ref{eqn:EOSlinder}, \ref{eqn:EOSparam}) relate $w(z)$ and its derivative, at low red shifts, to the equation of state in the large red shift limit (see Table \ref{tab:EOSparam}).  Our constraints on the large red shift value of the equation of state, $\lim_{z \rightarrow \infty} w(z)$, agree closely with those of \cite{Rapetti}.  These results appear to be independent of parameterization; analyses with three different parameterizations agree that $w(z)$ at large red shifts is slightly less than zero.

Our uncertainties in the dark energy parameters are mostly in agreement with similar analyses that include CMB and SN Ia data.  Our $2 \sigma$ uncertainties in $w_0$ and $w_0+w_a=\lim_{z \rightarrow \infty} w(z)$ are approximately twice as large as the $68.3 \%$ constraints of \cite{Rapetti}, just as expected.  Also, our $2 \sigma$ constraints on $w_0$ are about twice as large as the $1 \sigma$ uncertainties of \cite{Hannestad}.  Their looser bounds on $w'(0)$ can be attributed to the fact that they use a four-parameter equation of state, and marginalize over parameters other than those reported.  Meanwhile, comparison of our results with \cite{UrosSdssLyaBias} provides an example of parameterization effects on the dark energy constraints.  Their $95.5\%$ upper and lower bounds on $w_0$ are nearly the same as our $2 \sigma$ bounds, and their $95.5\%$ lower bound on $w_a$ is somewhat tighter than ours.  However, our upper bound on $w_a$ is smaller than theirs by a factor of three, due to the fact that our best-fit model is much nearer to the boundary $w_0+w_a=0$ of parameter space.  Still, the boundary has some effect on their $95.5\%$, and $99.86\%$ probability bounds on $w_a$.  Although their $68.32\%$ upper and lower bounds are nearly the same, their $99.86\%$ upper bound is smaller than the lower bound by a factor of two.  Thus, it would be interesting to see their analysis repeated with parameterizaton (\ref{eqn:EOSparam}), which has a less restrictive boundary.  

The effects of systematic uncertainties can be estimated by comparing the analysis procedures used by the studies listed Table \ref{tab:current}, as well as the locations of the resulting contours in parameter space.  Ref. \cite{UrosSdssLyaBias} finds $68\%$ and $95\%$ contours centered near the $\Lambda$CDM model, while \cite{Riess04, Hannestad, Alam, Choudhury, Rapetti, corasaniti} as well as our analysis find contours that lie mostly in the $w_0<-1$, $w'(0)>0$ region.  Even when the analysis of \cite{UrosSdssLyaBias} is repeated without the Lyman-$\alpha$ and galaxy bias data sets, their preferred models are still close to the $\Lambda$CDM model \cite{AlexeyMC}.  Since both studies use the standard SN Ia likelihood, the difference between their result and ours must come from either their CMB likelihood function \cite{Slosar} or their galaxy power spectrum analysis \cite{UrosSdssLyaBias, UrosSdssGalaxyBias}.  Ref. \cite{Slosar} claims that the WMAP likelihood approximation is inaccurate at low $l$, and that the WMAP foreground removal procedure may lead to a suppression of low-$l$ power.  Their likelihood function is designed to correct these problems.  Meanwhile, their galaxy power spectrum analysis uses SDSS measurements up to $k=0.2 h$/Mpc, a range which extends into the nonlinear regime.  The combined effect of these two changes is to move the contours very close to the $\Lambda$CDM model, which is disfavored by $\Delta \chi^2=3$ in our analysis with the same $w(z)$ parameterization.  Thus, differences in the CMB and galaxy power spectrum likelihood functions lead to a shift of over $1 \sigma$ in the $\chi^2$ contours.

\begin{table}[tb]   
\begin{center}
\begin{tabular}{|c||c|c|c|}
\hline
Analysis  &  Param.               & $w_0$ (to $2 \sigma$)   & $w_1$ or $w_a$ (to $2 \sigma$) \\
\hline
\hline
standard  & (\ref{eqn:EOSparam})  & $-1.38^{+0.30}_{-0.55}$ & $w_1=1.2^{+0.64}_{-1.06}$ \\
\hline
standard  & (\ref{eqn:EOSlinder}) & $-1.3^{+0.39}_{-0.34}$  & $w_a=1.25^{+0.40}_{-2.17}$ \\
\hline
CMB $l \geq 20$
          & (\ref{eqn:EOSparam})  & $-1.63^{+0.43}_{-0.65}$ & $w_1=1.5^{+0.45}_{-1.67}$ \\

\hline
$w_0+2w_1\geq 0$, 
          & (\ref{eqn:EOSparam})  & $-1.25^{+0.40}_{-0.38}$ & $w_1=0.6^{+0.23}_{-1.4}$, \\
CMB $l \geq 20$
          &                       &                         & $w_a \approx 2{\bar{w}}'=1.2^{+0.46}_{-2.8}$ \\
\hline
\end{tabular} 
\end{center}
\caption{ Consistency checks of our current $w(z)$ constraints.  All constraints shown are at the $2 \sigma$ level.  The ``standard'' analyses above use WMAP, SDSS, and SN Ia gold set data, with priors as shown in Table \ref{tab:priors}. \label{tab:testsOfCurrent}}
\end{table}  

We attempted to reproduce this shift by using only the multipoles $l \geq 20$ in our CMB analysis.  If the low multipoles were responsible for the shift, then neglecting them would enlarge the $\chi^2$ contours and move them towards the $\Lambda$CDM model.   However, as shown in the third column of Table \ref{tab:testsOfCurrent}, the allowed region actually moves away from that of \cite{UrosSdssLyaBias}, although it does broaden somewhat.  Not only does this test fail to explain the difference between the regions of parameter space preferred by \cite{UrosSdssLyaBias} and our results, it also demonstrates that the CMB analysis is sensitive to the low $l$ region of the power spectrum.  Since there is an ongoing debate about the proper handling of the low multipoles \cite{Slosar}, this sensitivity to that region of the power spectrum is worrisome.

Recall that switching dark energy parameterizations can change the uncertainties in $w(0)$ and $w'(0)$.  In order to test the effects of this parameterization dependence in combination with the low $l$ sensitivity, we imposed the prior constraint $w_0+2w_1=w_0+2{\bar{w}}' \geq 0$ on the analysis with CMB $l \geq 20$.  This allowed us to estimate the constraint on $w_a$ that we would have obtained if we had repeated the analysis with parameterization (\ref{eqn:EOSlinder}).  The resulting $2 \sigma$ constraints, shown in the fourth row of Table \ref{tab:testsOfCurrent}, are very similar to those found by the standard analysis using parameterization (\ref{eqn:EOSlinder}).  They have not moved appreciably towards the $\Lambda$CDM model, though the uncertainties have grown.  The slight shift away from $\Lambda$CDM seen without the prior $w_0+2{\bar{w}}' \geq 0$ has been hidden by the imposition of that prior.  This implies that the parameterization dependence of the results conceals some of their sensitivity to low $l$ CMB data.  Once again, we are unable to reproduce the shift in the location of the equation of state contours.  The explanation may have to do with the details of the CMB likelihood function used by \cite{UrosSdssLyaBias}, or with their galaxy power spectrum analysis.  These differences between the likelihood functions should be examined critically.

Similarly, the SN Ia likelihood function used by \cite{WangTegmark} causes the dark energy constraints to shift relative to those in \cite{Riess04}.  The standard supernova analysis \cite{Riess04} predicts $-1.59<w_0<-1.09$ and $0.58<w'<2.29$.  Using the same dark energy parameterization and the same prior $\Omega_m=0.27 \pm 0.04$, \cite{WangTegmark} obtains $-1.39<w_0<-0.25$ and $-2.61<w'<1.49$.  The \cite{WangTegmark} analysis uses supernova flux averaging \cite{Wang98} to handle systematic effects due to the weak lensing of supernovae \cite{Wang04}.  The flux averaging method assumes that uncertainties in SN Ia fluxes, rather than in magnitudes, are Gaussian.  If uncertainties are actually Gaussian in magnitude, this introduces a bias \cite{Wang98}; conversely, if uncertainties are Gaussian in flux, then the standard SN Ia analysis is biased.  The net effect of the flux averaged likelihood function is to weaken constraints on $w_0$ and $w'$, and to shift them both by about $1 \sigma$ towards the $\Lambda$CDM model.

Another SN Ia systematic effect, not included in any of the analyses in Table \ref{tab:current}, is the dimming of supernovae by astrophysical dust.  Ref. \cite{Ostman04} considers several types of intergalactic dust, and constrains supernova dust dimming to be less than $0.2$ magnitudes.  (For an extensive discussion of SN Ia systematics and their effects on the cosmological parameters, see \cite{Kim04}.)  The upper bound on dust dimming is not much less than the mean SN Ia magnitude uncertainty in the current data set, which suggests that dust dimming can cause a significant bias in dark energy constraints.  Since the dimming of SNe Ia provides evidence for accelerating cosmological expansion, additional dimming due to dust will appear to exaggerate this acceleration.  Ignoring systematic uncertanties due to dust dimming will cause an analysis to underestimate $w$, since more negative values of the equation of state lead to a more rapid acceleration. 

To summarize, current data are not precise enough to address whether or not the dark energy is a cosmological constant.  A careful study of the current data, and comparisons with other recent analyses, reveal several obstacles to a satisfactory understanding of the dark energy.  The degeneracy between $w_0$ and $w_1$ remains unresolved, and the allowed $2 \sigma$ ranges for $w_0$ and $w_1$ are at least $-1.93<w_0<-1.08$ and $0.14<w_1<1.84$.  Furthermore, this result is dependent on the parameterization chosen for the dark energy equation of state.  In particular, problems arise when the $1 \sigma$ and $2 \sigma$ contours lie near a boundary of the dark energy parameter space, as they do at present.  More generally, uncertainties in the equation of state parameters depends on the location of the best-fit model in the parameter space.  Aside from the parameterization dependence issues, analyses that use the same data, but different treatments of the systematics, find best-fit models differing by $\sim 1 \sigma$.  The combination of parameterization dependence and systematic effects is enough to shift the best-fit model by more than $2 \sigma$.  Thus we cannot rule out the $\Lambda$CDM model at present.


\subsection{Forecasts using Simulated Data}
\label{subsec:forecasts}

Next we ask, when data from future probes are available, how close we can come to achieving our goal of distinguising between a cosmological constant and other forms of dark energy.  The dark energy parameter space is continuous, with nonconstant $w(z)$ models arbitrarily close to the $\Lambda$CDM model.  Unless we are lucky and find a dark energy significantly different from the cosmological constant, the best we can do is to distinguish between models separated by some minimum distance in parameter space.  

Thus, we begin by identifying a set of reasonable milestones against which progress in $w(z)$ constraints may be measured.  Since we do not know ahead of time which best-fit model will be found, we will say that an experiment can distinguish between two points $A$ and $B$ in parameter space if, regardless of the location of the best-fit model, the $2 \sigma$ contour around that best-fit model excludes either $A$ or $B$ (or both).  

Our first milestone is to distinguish between a cosmological constant and a ``dark energy'' whose equation of state at $z=1$ is $w(1)=0$.  This goal is motivated by current observations, rather than by theoretical considerations; several of the analyses listed in Table \ref{tab:current}, including our own, favor $w \approx 0$ at red shifts of order unity.  We can distinguish between these two types of dark energy by ruling out either $w(1)=-1$ ($\Lambda$CDM model) or $w(1)=0$.  In the worst case, the best-fit equation of state will have $w(1)=-0.5$, meaning that the $2 \sigma$ uncertainty in $w(1)$ must be brought below $0.5$ in order to rule out one of the two types of dark energy.  Thus, the first milestone is decrease the uncertainty in $w$ at $z=1$ to less than $0.5$.

As our second milestone, we would like to distinguish between the $\Lambda$CDM model and tracker quintessence \cite{Zlatev98} or tracker SUGRA \cite{BraxMartin99} models with $w_0 \gtrsim -0.8$, $w_1>0$.  These are interesting, partly because they represent a large class of well established, theoretically motivated models, and also because they make definite predictions that are clearly distinct from the $\Lambda$CDM model in parameter space.  It is not clear whether such models are ruled out by current data; our analyses exclude them to greater than $2 \sigma$, but they are within the $95\%$ probability contour of \cite{UrosSdssLyaBias}.  In the future, we can compare an arbitrary best-fit model to $\Lambda$CDM and tracker models by comparing their values of $w(z_*)$, where $z_*$ is the red shift at which the uncertainty in $w(z)$ is a minimum.  The $\Lambda$CDM model predicts $w(z_*)=-1$, while trackers have $w(z_*) \gtrsim -0.8$.  In the worst case, the future best-fit model will have $w(z_*) \approx -0.9$.  Therefore, in order to reach the second milestone, we must reduce the $2 \sigma$ uncertainty in $w(z_*)$ to about $0.1$.

Finally, our third milestone is to distinguish between $\Lambda$CDM and quintessence models that are near to it.  We do not have in mind any particular class of theoretical models, so ``near'' is not well-defined.  As a reasonable third milestone, we consider aiming to distinguish $\Lambda$CDM from models with either $\left| w_0 + 1 \right| \gtrsim 0.05$ and $w_1=0$, or $w_0=-1$ and $\left| w_1 \right| \gtrsim 0.05$.  For this we will need both $\Delta w_0 (2 \sigma ) \lesssim 0.025$ and $\Delta w_1 (2 \sigma ) \lesssim 0.025$, that is, $\sigma_{w_0} \sim \sigma_{w_1} \sim 0.01$.  If the $\Lambda$CDM model were to be the preferred model even when the equation of state uncertainties were reduced to this level, then it would probably be time to abandon our hopes of using $w(z)$ to study the nature of dark energy.

Now that Sec. \ref{subsec:current} has pointed out several pitfalls in the analyses of current data, we make the resonably optimistic assumption that the likelihood functions associated with the CMB, SNe Ia, and cosmic shear will be well understood by the end of the decade.  We assume that these new, accurate likelihood functions will give no less cosmological information than the likelihood functions discussed in Sections \ref{subsec:snMC}, \ref{subsec:cmbMC}, and \ref{subsec:wlMC}.  This assumption is optimistic; the conservative CMB foreground removal procedure of \cite{Slosar}, as well as the inclusion of SN Ia dust dimming effects, should increase uncertainties in the dark energy parameters. Moreover, for weak lensing, several systematic effects ({\em{e.g.}} intrinsic alignments of source galaxies, selection biases and residuals from the PSF correction) need to be better understood and tightly controlled in order for this probe to achieve its full potential \cite{Refregier2}.

Also, our forecasts are based on a $\Lambda$CDM fiducial model, since ($w_0=-1$, $w'(0)=0$) is far from both the parameter space boundaries $w_0+w_1=0$ and $w_0+w_a=0$.  From now on, we limit our study to the parameterization (\ref{eqn:EOSparam}) alone.  Assuming that our $\chi^2$ contours stay away from the boundaries, we can compare constraints between parameterizations (\ref{eqn:EOSparam}) and (\ref{eqn:EOSlinder}) using the ``rule of thumb'' $\sigma_{w_1} = \sigma_{{\bar{w}}'} \approx \sigma_{w_a}/2$, where the average low red shift derivative ${\bar{w}}' \equiv w(1)-w(0)$.  The factor of $1/2$ means that, even when parameter space boundaries are unimportant, the uncertainty in the derivative $w'(0)$ is parameterization dependent.

\begin{figure}[tb]
\begin{center}
\includegraphics[width=3.3in]{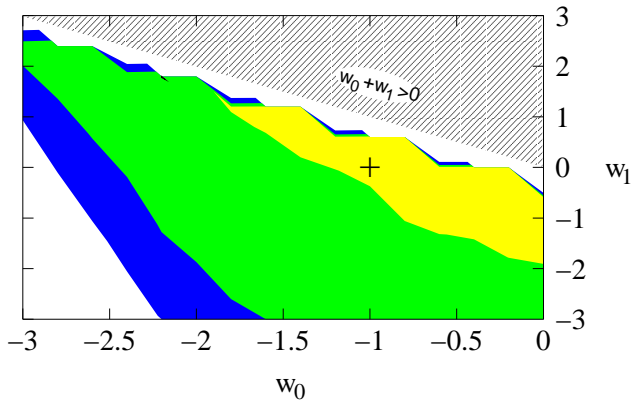}
\caption{Forecast CMB constraints on dark energy.  The $1\sigma$, $2\sigma$, and $3\sigma$ contours are shown.  The location of the fiducial model is marked by a "+" sign.}
\label{fig:fore-cmb}
\end{center}
\end{figure}

Dark energy constraints provided by eight years of simulated WMAP data alone are shown in Fig. \ref{fig:fore-cmb}, with grid spacings $\Delta_{w_0}=0.3$ and $\Delta_{w_1}=0.6$.  Models in the upper right-hand section of Fig. \ref{fig:fore-cmb} are ruled out by the prior constraint $w_0+w_1 \leq 0$.  Meanwhile, the requirement that $h<1.1$ eliminates models in the lower left-hand corner of the plot.  Therefore, to $2 \sigma$, virtually no constraints are imposed on $w_0$ and $w_1$ within the ranges shown in Fig. \ref{fig:fore-cmb}.  Such weak CMB constraints on dynamical dark energy are to be expected, given the angular diameter distance degeneracy discussed in Sec. \ref{subsec:cosmoParamDegenPriors}.

\begin{figure}[tb]
\begin{center}
\includegraphics[width=3.3in]{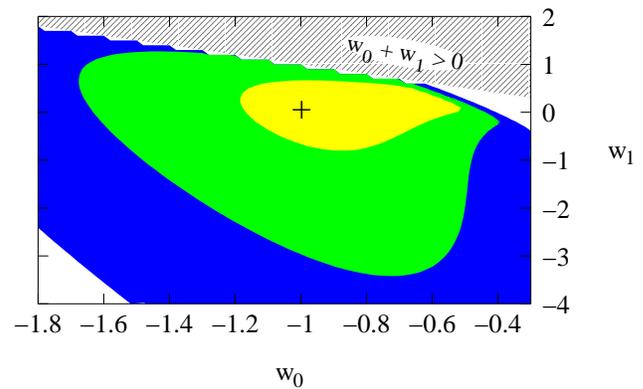}
\caption{SN Ia contours for the simulated dataset shown in Fig. \ref{fig:snDataSet}. The $1\sigma$, $2\sigma$, and $3\sigma$ contours are shown.   The location of the fiducial model is marked by a "+" sign.}
\label{fig:fore-sn_weak}
\end{center}
\end{figure}

Given the fairly weak priors used here, the supernovae alone were also unable to provide interesting constraints on dynamical dark energy.   Analysis of the supernova dataset shown in Fig. \ref{fig:snDataSet} gave deceptively tight constraints on $w_1$; see Fig. \ref{fig:fore-sn_weak}, with grid spacings $\Delta_{w_0}=0.02$ and $\Delta_{w_1}=0.1$.  However, multiple Monte Carlo simulations of SN Ia datasets yielded $2 \sigma$ constraints that varied widely from simulation to simulation.  Three out of our five simulations had $2 \sigma$ contours extending into the $w_1<-8$ range. 

The problem, as pointed out in \cite{PitfallsProspects}, is that the dark energy density $\rho_{de}(z)$ exhibits qualitatively very different behaviors for $w_1>0$ and $w_1<0$.  When $w_1>0$, $\rho_{de}(z)$ can remain a nontrivial fraction of the total energy density of the universe up to red shifts of order unity.  When $w_1<0$, $\rho_{de}(z)$ drops quickly with increasing red shift, so that the dark energy is important only in the very recent past.  Distinguishing between different dark energy models is difficult when $\rho_{de}(z)$ is very small.  This explains the fact that several of the simulated $2 \sigma$ contours extend deep into the $w_1<0$ region even though they are all simulations of a model with $w_1=0$.  

\begin{figure}[tb]
\begin{center}
\includegraphics[width=3.3in]{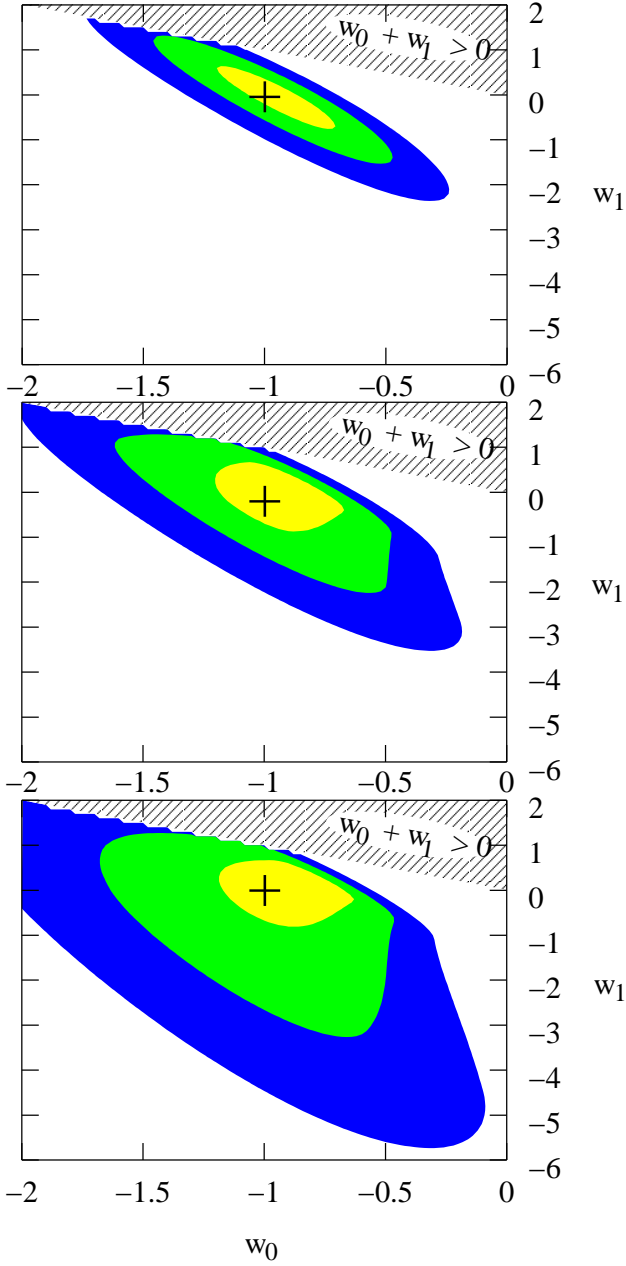}
\caption{SN Ia contours for the simulated dataset shown in Fig. \ref{fig:snDataSet}, assuming the prior constraints $\Omega_m=0.31$ (top), $\Omega_m=0.31 \pm 0.05$ (mid), and $\Omega_m=0.31 \pm 0.10$ (bot).  The $1\sigma$, $2\sigma$, and $3\sigma$ contours are shown.   The location of the fiducial model is marked by a "+" sign.}
\label{fig:fore-sn_strong}
\end{center}
\end{figure}

Adopting strong priors on $\Omega_m$ significantly reduces these non-Gaussianities, as shown in Fig. \ref{fig:fore-sn_strong}(top).  When we fix $\Omega_m=0.31$ and repeat the analysis of five simulated data sets, we find mean $2 \sigma$ uncertainties on $w_0$ and $w_1$ of $\Delta w_0 (2 \sigma) = 0.522$ and $\Delta w_1 (2 \sigma) = 1.63$.  The standard deviations in $\Delta w_0 (2 \sigma)$ and $\Delta w_1 (2 \sigma)$ are $0.036$ and $0.21$, respectively.  Thus when $\Omega_m$ is fixed, the uncertainties vary by only a few percent from one simulation to another.  

However, \cite{Virey} points out the perils of assuming strong priors in the SN Ia analysis.  In particular, if the analysis of another cosmological probe is based on the assumption that $w(z)=-1$, then the constraints from that analysis should not be imposed as priors in a study of dynamical dark energy.  Figs. \ref{fig:fore-sn_strong}(mid) and \ref{fig:fore-sn_strong}(bot) show the effects of weakening the prior constraint on $\Omega_m$ to $\Omega_m=0.31 \pm 0.05$ and $\Omega_m=0.31 \pm 0.10$, respectively.  Even with the relatively strong priors used in Fig. \ref{fig:fore-sn_strong}(mid), the non-Gaussianities have returned, and constraints on the dark energy parameters are weakened.  Thus it is imperative that the supernovae be combined with another data set in order to provide constraints on dynamical dark energy.

\begin{figure}[tb]
\begin{center}
\includegraphics[width=3.3in]{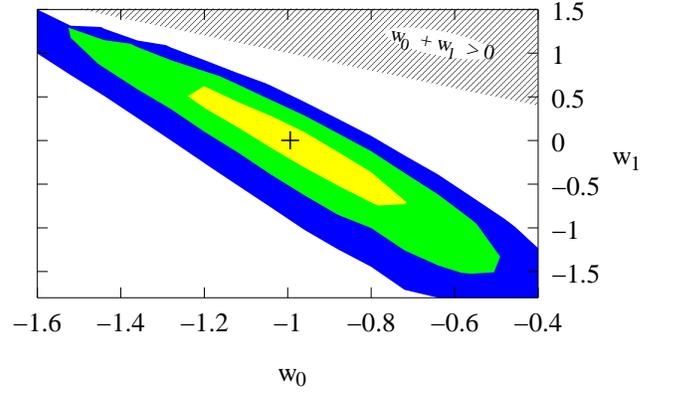}
\caption{Forecast CMB and SN Ia constraints on dark energy.  The $1\sigma$, $2\sigma$, and $3\sigma$ contours are shown.   The location of the fiducial model is marked by a "+" sign.}
\label{fig:fore-cmbSn}
\end{center}
\end{figure}

Combination of the WMAP-8 and SN Ia datasets improved dark energy constraints considerably, as shown in Fig. \ref{fig:fore-cmbSn}, with grid spacings $\Delta_{w_0}=0.08$ and $\Delta_{w_1}=0.22$. 
As with the supernova analysis, five simulated datasets were analyzed separately in order to determine the dark energy constraints.  We obtained the $2 \sigma$ constraints $\Delta w_0 (2 \sigma)=0.52$ and $\Delta w_1 (2 \sigma)=1.65$.  The standard deviations in the $2 \sigma$ $w_0$ and $w_1$ uncertainties were $0.027$ and $0.25$, respectively.  This implies that our forecast uncertainties are each correct to about $10\%$.  We checked these constraints using a Fisher matrix calculation and found the $2 \sigma$ constraints $0.44$ and $1.46$ in $w_0$ and $w_1$, respectively, consistent with our minimization results.  Meanwhile, the minimum $2 \sigma$ uncertainty in $w(z)$ was found to be $\Delta w(z_*) (2 \sigma )=0.19$ at $z_* = 0.27$.

Our forecast uncertainty, in each of the equation of state parameters, is greater by a factor of 2-3 than that of \cite{SNAP04} with SNe Ia and a prior on $\Omega_m$, as shown in Table \ref{tab:forecast}.  Note that our $1 \sigma$ uncertainties have been computed by dividing our $2 \sigma$ uncertainties by two.

\begin{table}[!tb]  
\begin{center}
\begin{tabular}{|l||c|l|c|}
\hline
Ref.             & surveys         & parameters varied;            & forecast \\
                 & included        & priors                        & constraints \\
\hline
\hline 
\cite{Hu01}      & Planck + WL     & ``vanilla'', $w_0$, $T/S$;    & $\sigma_{w_0}=0.15$ \\
                 & (1000 $deg^2$)  & $w'=0$                        & \\
\hline
\cite{Huterer01} & WL + COBE       & $\omega_m$, $\omega_b$, $n_s$, $m_{\nu}$, $w_0$;
                                                                   & $\sigma_{w_0}=0.19$ \\
                 & + photo z       & $w'=0$,                       & \\
                 & + Planck        & $\sigma_{\ln \omega_m}=.064$, & \\ 
                 &                 & $\sigma_{\ln \omega_b}=.035$, & \\
                 &                 & $\sigma_n=.04$,               & \\
		 &                 & $\sigma_{m_{\nu}}=.58$        & \\
\hline
\cite{SNAP04}    & SNAP SNe Ia     & $w_0$, $w_a$, $\Omega_m$;     & $\sigma_{w_0}=0.09,$ \\
                 & (syst. $\delta m = $ 
                                   & $\sigma_{\Omega_m}=0.03$,     & $\sigma_{\bar{w}'}=0.31$ \\
                 & $0.0074 (1+z)$)                & $\Omega_K=0$                  & \\
\hline 
\cite{SNAP04}    & SNAP SNe Ia     & $w_0$, $w_a$, $\Omega_m$;     & $\sigma_{w_0}=0.09,$ \\
                 & (syst. $\delta m = $
                                   & Planck priors,                & $\sigma_{\bar{w}'}=0.19$ \\
                 & $0.0074 (1+z)$) & $\Omega_K=0$                  & \\
\hline 
\cite{SNAP04}    & SNAP SNe Ia     & $w_0$, $w_a$, $\Omega_m$;     & $\sigma_{w_0}=0.05,$ \\
                 & (syst. $\delta m = $
                                   & $\Omega_K=0$                  & $\sigma_{\bar{w}'}=0.11$ \\
                 & $0.0074 (1+z)$) &                               & \\
                 & + WL            &                               & \\
                 & ($f_{sky}=0.025$) &                             & \\
\hline 
\cite{SongKnox}  & Planck + WL     & ``vanilla'', $w_0$, $w_a$,    & $\sigma_{w_0}=0.056$, \\
                 & ($f_{sky}=1$,   & $\omega_{\nu}$, $\alpha_s$, $y_{He}$  
		                                                   & $\sigma_{w_a}=0.087$\\
	   	 & tomography)	   &                               & \\  
\hline
\cite{SongKnox}  & Planck + WL     & ``vanilla'', $w_0$, $w_a$,    & $\sigma_{w_0}=0.076$, \\
                 & ($f_{sky}=0.5$, & $\omega_{\nu}$, $\alpha_s$, $y_{He}$
		                                                   & $\sigma_{w_a}=0.11$\\
	   	 & tomography)	   &                               & \\  
\hline
\cite{SongKnox}  & 4 yr. WMAP +    & ``vanilla'', $w_0$, $w_a$,    & $\sigma_{w_0}=0.064$, \\
                 & WL ($f_{sky}=1$,& $\omega_{\nu}$, $\alpha_s$, $y_{He}$ 
		                                                   & $\sigma_{w_a}=0.11$\\
	   	 & tomography)	   &                               & \\ 
\hline
\cite{Albrecht}  & Planck +        & ``vanilla'', $w_0$, $w_a$,    & $\sigma_{w_0}=0.05$, \\
                 & SNAP SNe Ia +   & $\omega_{\nu}$, $\alpha_s$, $y_{He}$ 
		                                                   & $\sigma_{w_a}=0.1$\\
	   	 & WL ($f_{sky}=0.5$, &                            & \\   
		 & tomography)	   &                               & \\
\hline
\cite{WangClusterCounts04}
                 & Planck + LSST   & $\omega_b$, $\omega_m$, $\Omega_{de}$, $\sigma_8$, 
                                                                   & $\sigma_{w_0}=0.036$, \\
                 & cluster counts  & $n_s$, $w_0$, $w_a$;          & $\sigma_{w_a}=0.093$ \\
                 & and power       & linear biasing                & \\
                 & spectrum        &                               & \\
                 & (200,000 clusters) &                            & \\

\hline
\cite{LinderMiquel04}
                 & 1280 SNe \footnotemark[2] 
                   \footnotetext[2]{irreducible magnitude systematic $\delta m^{irr} = 0.01+0.06z$ for $z<0.9$ or $0.1z$ for $z>0.9$, 
                                    extinction correction uncertainty $\delta m^{ext} = 0.02$, 
				    low-mid $z$ magnitude offset uncertainty $\delta m^{l-m,off}=0.02$, 
				    mid-high $z$ magnitude offset uncertainty $\delta m^{m-h,off}=0.04$ }
                                   & $w_0$, $w_a$, $\Omega_m$;     & $\sigma_{w_0}=0.27,$ \\
                 &                 & Planck priors,          & $\sigma_{\bar{w}'}=0.57$ \\ 
                 &                 & $\sigma_{\Omega_m}=0.03$,     & \\
                 &                 & $\Omega_K=0$                  & \\

\hline
AU,              & 8 yr. WMAP +    & ``vanilla'', $w_0$, $w_1$;    & $\sigma_{w_0}=0.26,$ \\
MI,              & $\sim2000$ SNe  & for priors see                & $\sigma_{w_1}=0.82$ \\
PJS              & (syst. $\delta m=0.04$)                
                                   & Table \ref{tab:priors}        & \\
\hline
AU,              & 8 yr. WMAP +    & ``vanilla'', $w_0$, $w_1$;    & $\sigma_{w_0}=0.10,$ \\
MI,              & $\sim2000$ SNe  & for priors see                & $\sigma_{w_1}=0.18$ \\
PJS              & (syst. $\delta m=0.04$)                
                                   & Table \ref{tab:priors}        & \\
                 & + WL            &                               & \\
\hline
\end{tabular}
\end{center}
\caption{Forecast $1 \sigma$ constraints on the dark energy equation of state. (The ``vanilla'' parameter space is spanned by the six parameters ($h$, $\omega_m$, $\omega_b$, $\tau$, $A$, $n_s$).)\label{tab:forecast}}
\end{table}

\begin{figure}[tb]
\begin{center}
\includegraphics[width=3.3in]{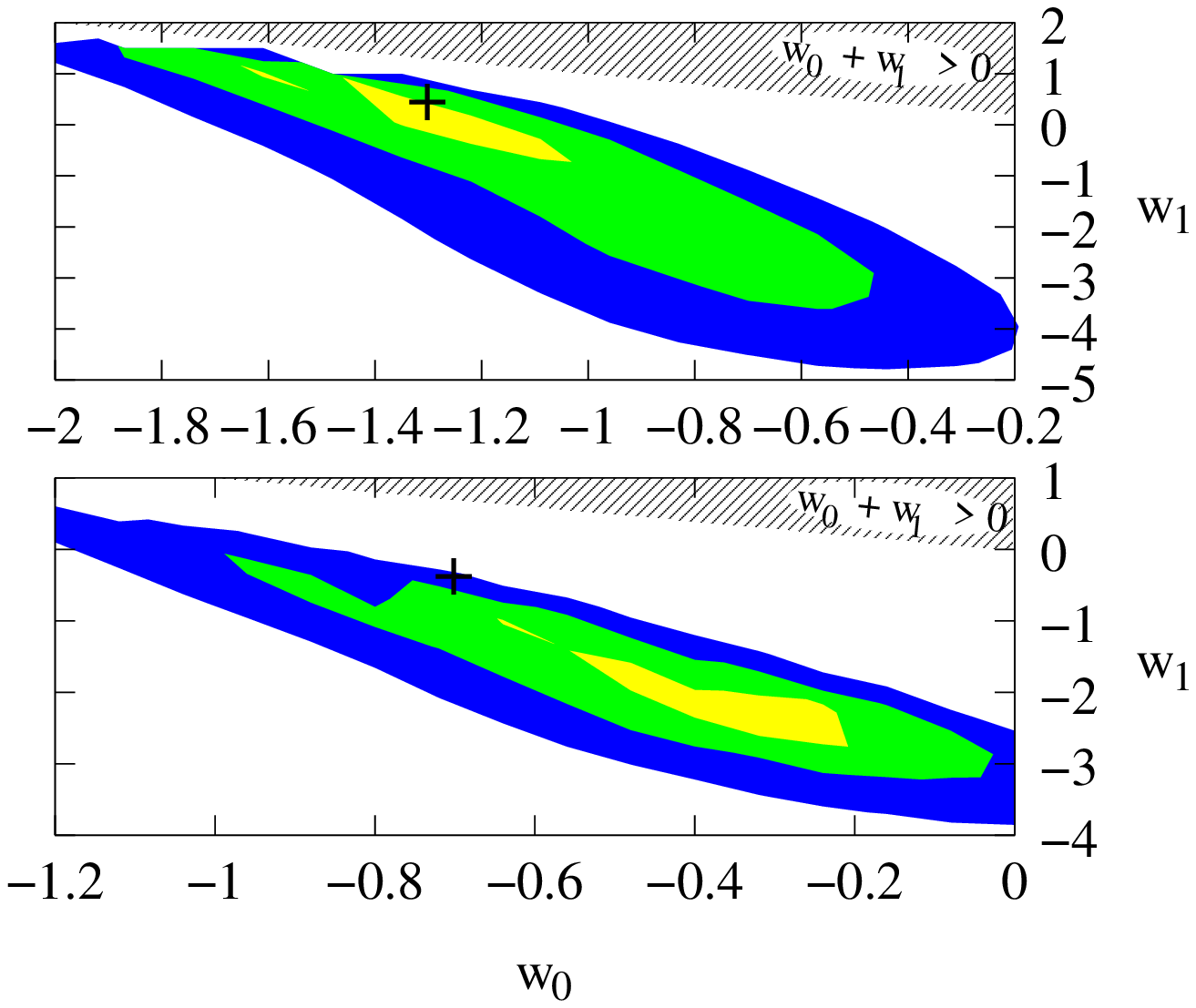}
\caption{$\chi^2$ contours for CMB and SN Ia ``data'' simulated using the fiducial models ($w_0=-1.3$, $w_1=0.5$) (top) and ($w_0=-0.7$, $w_1=-0.3$) (bottom). The $1\sigma$, $2\sigma$, and $3\sigma$ contours are shown.    The location of the fiducial model is marked by a "+" sign.}
\label{fig:cmbSn.varfid}
\end{center}
\end{figure}

\begin{table}[tb]  
\begin{center}
\begin{tabular}{|p{0.5in}||p{0.9in}|c|c|}
\hline
Probes       &    Analysis           & $\Delta w_0 (2 \sigma)$    & $\Delta w_1 (2 \sigma)$ \\
\hline
\hline
SN Ia        & prior $\Omega_m=0.31$ & $0.52 \pm 0.036$           & $1.63 \pm 0.21$ \\
\hline
CMB +        & standard              & $0.52 \pm 0.027$           & $1.65 \pm 0.25$ \\ \cline{2-4}
SN Ia        & $f_{sky}^{\phantom{sky}2}\rightarrow f_{sky}$ \newline in (\ref{eqn:dCTT}-\ref{eqn:CovTEEE})
                                     & $0.52$                     & $1.31$ \\ \cline{2-4}
             & CMB $l \leq 400$      & $0.50$                     & $1.42$ \\ \cline{2-4}
             & 1 yr. WMAP            & $0.47 \pm 0.09$            & $1.76 \pm 0.42$ \\ \cline{2-4}
             & SN $\delta m = 0.06$  & $0.65$                     & $1.75$ \\ \cline{2-4}
             & SN $\delta m = 0.06z$ & $0.31$                     & $0.997$ \\ \cline{2-4}
             & SN $\delta m = 0$     & $0.19 \pm 0.018$           & $0.68 \pm 0.071$ \\ \cline{2-4}
\hline
CMB +        & standard              & $0.20 \pm 0.021$           & $0.37 \pm 0.034$ \\ \cline{2-4}
SN Ia +      & WL on linear          & $0.39$                     & $0.96$ \\
WL           & scales only           &                            & \\ \cline{2-4}
             & CMB $l \leq 400$      & $0.23$                     & $0.47$ \\ \cline{2-4}
             & SN $\delta m = 0.06$  & $0.24$                     & $0.48$ \\ \cline{2-4}
             & SN $\delta m = 0$     & $0.12 \pm 0.007$           & $0.29 \pm 0.03$ \\ \cline{2-4}
\hline
\end{tabular}
\end{center}
\caption{Consistency checks of our dark energy constraints.  All constraints shown are at the $2 \sigma$ level, with the standard deviations in constraints from multiple Monte Carlo simulations shown when available.  The ``standard'' analyses are those described in Sections \ref{subsec:snMC}, \ref{subsec:cmbMC}, and \ref{subsec:wlMC}, while the other analyses differ from the standard ones as specified in the Analysis column.  \label{tab:checkFore}}
\end{table}

Next, the analysis was repeated with two different fiducial models, chosen to lie approximately along the line of degeneracy in the ($w_0$,$w_1$) plane.  As shown in Fig. \ref{fig:cmbSn.varfid}, the model ($w_0=-1.3$, $w_1=0.5$) had $2 \sigma$ uncertainties of 0.73 and 2.58 on $w_0$ and $w_1$, respectively, which are significantly larger than those reported above.  Meanwhile the model ($w_0=-0.7$, $w_1=-0.3$) had uncertainties of 0.48 and 1.58.  These are consistent with our constraints obtained using the $\Lambda$CDM fiducial model.

Returning to our $\Lambda$CDM fiducial model, we repeated the analysis with different assumptions about the data in order to test the robustness of our forecast constraints.  Table \ref{tab:checkFore} lists the new constraints obtained.  The standard deviations in our original constraints are about $10\%$, so we did not consider a modification to be significant unless it changed constraints by at least $20\%-30\%$.  First we checked to what extent a CMB simulation without the ``extra'' factor of $f_{sky}$ in the denominator of the covariance matrix (\ref{eqn:dCTT}, \ref{eqn:dCEE}, \ref{eqn:dCTE}, \ref{eqn:CovXX}, \ref{eqn:CovTTTE}, \ref{eqn:CovTTEE}, \ref{eqn:CovTEEE}) would improve our constraints.  The difference turns out to be negligible.  In fact, the final constraints are relatively independent of the details of the CMB simulation and analysis.  Cutting off the CMB power spectrum at a maximum multipole of $400$, or replacing the WMAP 8 year data set with a simulated 1 year data set, lead to insignificant changes in the dark energy constraints.  On the other hand, the constraints are very sensitive to changes in the quality of the supernova data.  Changing the SN Ia systematic uncertainty, or its red shift dependence, leads to significant changes in the dark energy constraints.  Meanwhile, the final $2 \sigma$ uncertainties provided by the combination of CMB and SN Ia data are nearly identical to those found by fixing $\Omega_m$ in the SN analysis alone.  Thus, in some sense, adding the CMB data set is equivalent to fixing $\Omega_m$ as a function of $w_0$ and $w_1$ in the supernova analysis.

We began Sec. \ref{subsec:forecasts} by identifying three milestones for comparing constraints on $w_0$ and $w_1$.  Recall that the first milestone is to distinguish between a cosmological constant and a dark energy with $w(1)=0$.  From our forecasts we find $\Delta w(1) (2 \sigma) = 1.11$, which is more than twice as high as the uncertainty of $0.5$ needed to distinguish between the two dark energy models.  The second milestone, to distinguish between $\Lambda$CDM and tracker models of dark energy, is also not yet reached by the combination of CMB and SN Ia data sets.  We find $\Delta w(z_*) (2 \sigma) = 0.19$ at $z_* = 0.27$, twice the uncertainty of $0.1$ needed to reach this milestone.  We can see from Fig. \ref{fig:fore-cmbSn} that, if we shifted the contours to be centered on ($w_0=-0.9$, $w_1=0$), then the cosmological constant model as well as some of the tracker models ($w_0 \geq -0.8$, $w_1>0$) would be within the $2 \sigma$ contours.  Thus we cannot confidently claim that the combination of CMB and SN Ia data can rule out either the $\Lambda$CDM model or the trackers.  Finally, this combination is far from reaching the third milestone, which would require that the uncertainties in $w_0$ and $w_1$ be reduced by over an order of magnitude.



\begin{figure}[tb]
\begin{center}
\includegraphics[width=3.3in]{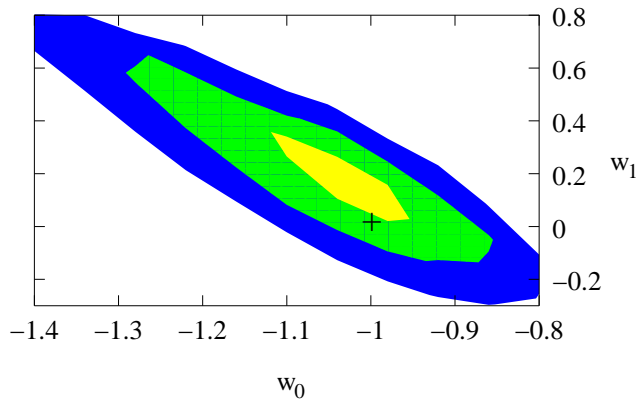}
\caption{Forecast CMB, SN Ia, and WL constraints on dark energy.  The $1\sigma$, $2\sigma$, and $3\sigma$ contours are shown.   The location of the fiducial model is marked by a "+" sign.}
\label{fig:fore-cmbSnWl}
\end{center}
\end{figure}

The addition of weak lensing to the analysis resulted in the contours shown in Fig. \ref{fig:fore-cmbSnWl}, with grid spacings $\Delta_{w_0}=0.06$ and $\Delta_{w_1}=0.12$.  Weak lensing tightened the $2 \sigma$ constraints to $\Delta w_0 (2 \sigma)=0.20$ and $\Delta w_1 (2 \sigma)=0.37$.  The standard deviations in the $2 \sigma$ uncertainties, found from five separate simulations, were $0.021$ and $0.034$ in $\Delta w_0 (2 \sigma)$ and $\Delta w_1 (2 \sigma)$, respectively.  As above, the standard deviations in each of the $w_0$ and $w_1$ uncertainties were about $10\%$.
 We checked our results using a Fisher matrix calculation and found the $2 \sigma$ constraints $0.16$ and $0.38$ in $w_0$ and $w_1$, respetively, consistent with our minimization results.
Meanwhile, the minimum $2 \sigma$ uncertaity in $w(z)$ was found to be $\Delta w(z_*) (2 \sigma) = 0.097$ at $z_* = 0.52$.

  Our constraints are about the same as those forecast for SNAP SNe Ia with Planck priors, but worse by a factor of two than SNAP SNe Ia with weak lensing, as listed in Table \ref{tab:forecast}.  Meanwhile, galaxy cluster measurements based on the more ambitious LSST survey claim an improvement by a factor of three over our forecast $w_0$ constraints \cite{WangClusterCounts04}.

\begin{figure}[tb]
\begin{center}
\includegraphics[width=3.3in]{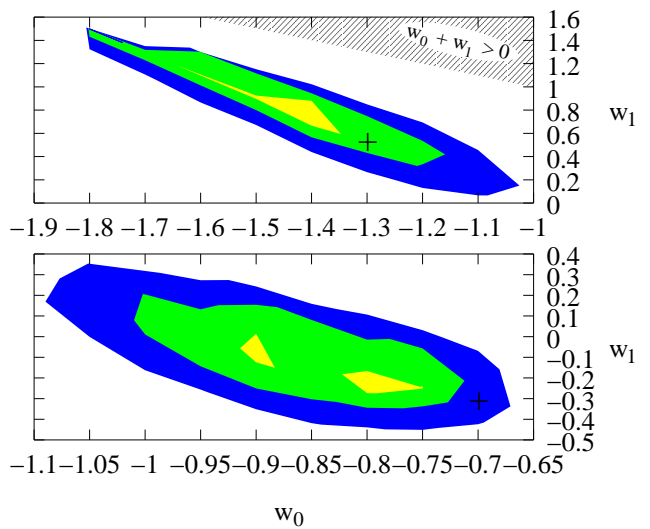}
\caption{$\chi^2$ contours for CMB, SN Ia, and WL ``data'' simulated using the fiducial models ($w_0=-1.3$, $w_1=0.5$) (top) and ($w_0=-0.7$, $w_1=-0.3$) (bottom). The $1\sigma$, $2\sigma$, and $3\sigma$ contours are shown.    The location of the fiducial model is marked by a "+" sign.}
\label{fig:cmbSnWl.varfid}
\end{center}
\end{figure}

The analysis was repeated with two different fiducial models in order to assess the dependence of dark energy constraints on the fiducial model.  Moving from the $\Lambda$CDM fiducial to the fiducial model ($w_0=-1.3$, $w_1=0.5$) increases uncertainties in $w_0$ and $w_1$ by about $70 \%$ each, to $0.34$ and $0.63$, respectively.  Moving in the other direction along the degeneracy curve, to the fiducial model ($w_0=-0.7$, $w_1=-0.3$), leads to a modest decrease in $w_0$ and $w_1$ uncertainties to $0.15$ and $0.28$, respectively.  Qualitatively, this is the same behavior as was seen with the CMB and SN Ia data combination.

Further tests of the robustness of our constraints revealed that, once again, CMB data at $l>400$ do not contribute much to the dark energy constraints (see Table \ref{tab:checkFore}). Meanwhile, if the weak lensing analysis is restricted to linear scales, then constraints on $w_0$ and $w_1$ weaken considerably, as shown in Table \ref{tab:checkFore}.  This is consistent with the findings of \cite{Huterer01}.

\begin{figure}[tb]
\begin{center}
\includegraphics[width=2.3in, angle=270]{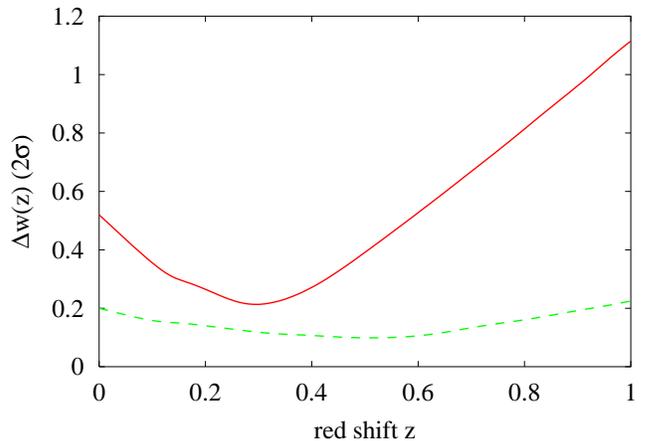}
\caption{$\Delta w(z) (2 \sigma)$ as a function of $z$, for the data combinations CMB + SN Ia (upper, solid line) and CMB + SN Ia + WL (lower, dashed line).}
\label{fig:2_sigma_w_vs_z}
\end{center}
\end{figure}

Comparing $2 \sigma$ constraints with and without weak lensing (including nonlinear scales), we see that the uncertainties in $w_0$ shrink by a factor of $2.5$, and the uncertainties on $w_1$ shrink by a factor of $4.5$.  In order to understand the contribution of weak lensing to the overall analysis, we compared $w(z)$ uncertainties as functions of redshift for the combinations CMB + SN Ia and CMB + SN Ia + WL, as shown in Fig. \ref{fig:2_sigma_w_vs_z}.  Without weak lensing, $w(z)$ is well constrained only around $z \approx 0.3$.  Evidently, weak lensing adds information on $w$ at some higher red shift, complementing the constraints from CMB and SN Ia.  The result is a $w(z)$ uncertainty that is not only lower, but much more uniform across the red shift range.

This improvement allows the combination of CMB, SN Ia, and WL data to reach two of the three milestones identified at the beginning of this section.  At $z=1$ we find $\Delta w(1) (2 \sigma) = 0.22$, which will easily allow us to distinguish between a cosmological constant and dark energy models with $w(1) \approx 0$.  Thus, weak lensing will either confirm or conclusively rule out the dark energies with $w_0 + w_1 \lesssim 0$ favored by our analysis of current data.  Secondly, the constraint $\Delta w(z_*) (2 \sigma ) = 0.097$ at $z_* = 0.52$ is tight enough that this combination of data can distinguish between the cosmological constant and tracker models of dark energy.  Thus, weak lensing will allow us to rule out a portion of the most interesting region of parameter space.  On the other hand, our forecast dark energy constraints do not reach the third milestone, which calls for these $2 \sigma$ uncertainties to be reduced to $\sim 0.025$.  In particular, the uncertainty in $w_1$ is higher than this by over an order of magnitude.

We have not included in our analysis either weak lensing tomography or galaxy cluster measurements, which may lower statistical uncertainties even further \cite{IshakSpergel04, WangClusterCounts04}.  These have been analyzed by others and look promising.  They suggest that we can approach the third milestone, but there is no method suggested so far for pushing substantially beyond their forecast constraints.

Moreover, we remind the reader that the constraints discussed above, as well as the improvement due to weak lensing, are based on several optimistic assumptions listed at the beginning of Sec. \ref{subsec:forecasts}.  Systematic effects, such as dust-related dimming of the supernovae, can increase uncertainties in $w_0$ and $w_1$.  Also, if the best-fit dark energy model is near the boundary of the ($w_0$, $w_1$) parameter space, then $w(z)$ parameterization dependence will further weaken our constraints on the dark energy parameters.  In the worst-case scenario $w_0+w_1 \approx 0$, the best fit model will be near the boundary of parameter space; parameterization-related uncertainties in $w_0$ and $w'(0)$ could completely swamp any improvements.  


\section{Conclusions}
\label{sec:conclusions}

The purpose of this investigation has been to determine how well $w(z)$ can be resolved using currently planned astronomical observations.  The well-known challenge is that individual measurements do not constrain $w(z)$ directly, but rather some functional that depends on $w(z)$, integrals of $w(z)$, and a large set of additional parameters.  

We have shown that the uncertainties in measuring $w$ and $dw/dz$ at $z=0$ can be reduced dramatically by the beginning of the next decade, using a combination of the highest-quality CMB, SN, and WL data.  However, the remaining uncertainties, $\Delta w_0 (2 \sigma ) = 0.20$ and $\Delta w_1 (2 \sigma )=0.37$, will not be enough to determine definitively whether dark energy is inert (a cosmological constant) or dynamical (quintessence), unless the true value of $w$ differs from $-1$ by significantly more than $0.1$.  Unfortunately, many quintessence models have $\left| w_0+1 \right| <0.1$ and $\left| w_1 \right| <0.1$.

Our numerical studies illustrate how the measurements combine to produce this constraint.  Even the best supernova measurements are degenerate under certain combinations of variations in $w(0)$, $w'(0)$, and $\Omega_m$ \cite{Maor2001}.  CMB constraints are degenerate along a surface in the space spanned by $w(0)$, $w'(0)$, $h$, and $\Omega_m$.  However, we have found (Fig. \ref{fig:fore-cmbSn}) that combining CMB and SN measurements effectively collapses the degeneracy in the $\Omega_m$ direction, leaving only the degeneracy in the $w(0)$, $w'(0)$ plane. We have shown that the CMB contribution to this degeneracy breaking comes entirely from the $l \leq 400$ region of the power spectrum.  Since even the first year of WMAP data has reduced uncertainties in this range to near the cosmic variance level, we do not expect a significant improvement in dark energy constraints from further WMAP or Planck CMB data.  Of course, this relies on our choice and number of cosmological parameters, and in particular, our assumption that $\Omega_K=0$.  If curvature were considered, the angular diameter distance degeneracy would degrade the information on $\Omega_m$ provided by the first CMB acoustic peak, and better information on subsequent peaks could prove valuable.  Also, we have not considered nonlinear effects such as the gravitational lensing of the CMB power spectra \cite{ChrisUrosCMBlensing03}, which could provide valuable information on structure formation.  

Meanwhile, the final constraints are sensitive to the supernova systematic uncertainties.  A $50\%$ increase in $\delta m$, from $0.04$ to $0.06$, changes the $2 \sigma$ constraints on $w(0)$ by about $25 \%$.  After the CMB and SN data sets are combined, the remaining degeneracy runs along curves of $w(z)$ which intersect one another near $z=0.3$.  The uncertainty in $w(0.3)$ is roughly $0.2$ at the $2 \sigma$ level, depending on what functional forms for $w(z)$ are considered. 

To break the degeneracy between $w(0)$ and $w'(0)$, more data must be co-added that can constrain $w(z)$ at a greater red shift.  We have studied the weak lensing power spectrum as a means of breaking the $w(0)$-$w'(0)$ degeneracy.  Our previous conclusion about the highest CMB multipoles remains valid; CMB data at $l>400$ do not contribute to dark energy constraints even when weak lensing is considered.  Furthermore, any improvements due to weak lensing depend crucially on our ability to use shear measurements on nonlinear scales.  This will require a better understanding of systematics, such as intrinsic alignments of galaxies, as well as more accurate computations of the matter power spectrum on nonlinear scales in dynamical dark energy cosmologies.  When CMB, SN, and WL are combined, the supernovae constrain $w(z)$ at low red shifts $z\approx 0.3$, while weak lensing constrains $w(z)$ at higher red shifts (see Fig. \ref{fig:2_sigma_w_vs_z}), leading to improvements of a factor of $2.5$ in $w_0$ and a factor of $4.5$ in $w_1$.  

Section \ref{subsec:forecasts} began by identifying three milestones by which progress in dark energy constraints could be measured.  
\begin{itemize}
  \item Distinguish between $w(1)=-1$ and $w(1) \approx 0$.  CMB and SN alone are unable to reach this milestone, while the combination of CMB, SN, and WL data can distinguish between these equations of state at the $4.5 \sigma$ level. 
  \item Distinguish between $w=-1$ and tracker models with $w_0 \geq -0.8$, $dw/dz > 0$.  Once again, CMB and SN alone are unable to reach this milestone.  With WL added, $\Lambda$ and trackers can be distinguished at the $2 \sigma$ level. 
  \item Reduce $2 \sigma$ uncertainties in $w$ and $dw/dz$ to $\approx 0.025$.  Even with CMB, SN, and WL data, this milestone remains unreached.
\end{itemize}
Thus the combination of CMB, SN Ia, and weak lensing is a promising tool for improving dark energy constraints.  However, in the worst-case scenario that experiments find $w \approx -1$, these three probes cannot decisively rule out quintessence models in which $w$ differs from $-1$ by a few percent.  

Other works have considered highly ambitious surveys of the cluster abundance evolution \cite{WangClusterCounts04}, assuming large numbers of observed clusters, or of weak lensing tomography \cite{IshakSpergel04}, based on multiple red shift bins and observations of large fractions of the sky.  Both probes measure the structure growth rate as a function of red shift.  These have the potential to improve statistical uncertainties significantly, by factors of 3-5 compared with our results, but concerns remain about systematic uncertainties \cite{IshakSpergel04, WangClusterCounts04}.  Even taking the current estimated errors, uncertainties in $w_0$ and $w_1$ are several percent, which still allows a range of plausible quintessence models.  Thus, unless we are lucky enough to find a dark energy that is very different from the cosmological constant, new kinds of measurements or an experiment more sophisticated than those yet conceived will be needed in order to settle the dark energy issue.


%
%
\acknowledgments
We thank N. Bahcall, R. Brustein, J. Gunn, C. Hirata, S. Ho, E. Linder, A. Makarov, I. Maor, J. Ostriker, A. Riess, U. Seljak, D. Spergel, C. Stubbs,  Y. Wang, and D. Wesley for useful 
conversations.
Our analysis was run on a Beowulf cluster at Princeton University,
supported in part by NSF grant AST-0216105.  
This material is based upon work supported under a National Science Foundation Graduate Research Fellowship (AU).
M.I. acknowledges the support of the Natural Sciences and Engineering Research Council of Canada.
This work was supported in part by US Department of Energy Grant DE-FG02-91ER40671 (PJS).

\end{document}